\documentclass[]{aa} 
\usepackage{txfonts}
\usepackage{natbib}
\usepackage{graphicx, xspace}

\usepackage[switch]{lineno}

\newcommand{\chisq} {$\chi$$^2$}
\newcommand{\Msun}{\>{M_{\odot}}}
\newcommand{\Rsun}{\mbox{$R_{\odot}$}}

\newcommand{\degrees} {^\circ}
\newcommand{\degs} {^\circ}

\makeatletter

\newcommand{\Rmnum}[1]{\expandafter\@slowromancap\romannumeral #1@}
\makeatother

\usepackage[dvips]{color}

\newcommand{\new}[1]{#1}
\newcommand{\newer}[1]{#1}
\renewcommand{\new}[1]{#1}

\def\dn1{$\delta\nu_{01}$}
\def\dn2{$\delta\nu_{02}$}
\def\sun{\hbox{$_\odot$}}

\begin{document}

\linenumbers
\modulolinenumbers[5]

\title{Pulsating red giant stars in
eccentric binary systems discovered from \textit{Kepler} space-based photometry
\subtitle{A sample study and the analysis of KIC\,5006817}
}
\authorrunning{P.\,G. Beck et al.}
\titlerunning{Pulsating red giant stars in eccentric binaries}

\author{P.~G.~Beck\inst{1}, 
K.~Hambleton\inst{2,1}, 
J.~Vos\inst{1},
T.~Kallinger\inst{3}, 
S.~Bloemen\inst{1}, 
A.~Tkachenko\inst{1},  
R.\,A.~Garc\'ia\inst{4}, 
R.\,H.~\O stensen\inst{1},
C.~Aerts\inst{1,5}, 
D.\,W.~Kurtz\inst{2}, 
J.~De\,Ridder\inst{1}, 
S.~Hekker\inst{6}, 
K.~Pavlovski\inst{7}, 
S.~Mathur\inst{8}, 
K.~De\,Smedt\inst{1},
A.~Derekas\inst{9},
E.~Corsaro\inst{1}, 
B.~Mosser\inst{10},
H.~Van\,Winckel\inst{1}, 
D.~Huber\inst{11},
P.~Degroote\inst{1}, 
G.\,R.~Davies\inst{12},
A.~Pr\v{s}a\inst{13},
J.~Debosscher\inst{1}, 
Y.~Elsworth\inst{12},
P.~Nemeth\inst{1}, 
L. Siess\inst{14},
V.\,S. Schmid\inst{1},
P.\,I.~P\'apics\inst{1}, 
B.\,L.~de\,Vries\inst{1}, 
A.\,J.~van\,Marle\inst{1},
P.\, Marcos-Arenal\inst{1},
A.~Lobel\inst{15}}

\date{Received 12 August 2013; Accepted 16 December 2013}


\institute{
Instituut voor Sterrenkunde, KU Leuven, 3001 Leuven, Belgium. 
\email{paul.beck@ster.kuleuven.be}
\and Jeremiah Horrocks Institute, University of Central Lancashire, Preston PR1 2HE, UK. 
\and Institut f\"ur Astronomie der Universit\"at Wien, T\"urkenschanzstr. 17, 1180 Wien, Austria. 
\and Laboratoire AIM, CEA/DSM Ð CNRS - Universit\'e Denis Diderot Ð IRFU/SAp, 91191 Gif-sur-Yvette Cedex, France. 
\and Department of Astrophysics, IMAPP, University of Nijmegen, PO Box 9010, 6500 GL Nijmegen, The Netherlands\label{inst2} 
\and Astronomical Institute ÕAnton PannekoekÕ, University of Amsterdam, Science Park 904, 1098 XH Amsterdam, The Netherlands.
\and Department of Physics, Faculty of Science, University of Zagreb, Croatia. 
\and Space Science Institute, 4750 Walnut street Suite\#205, Boulder, CO 80301, USA 
\and Konkoly Observ., Research Centre f. Astronomy and Earth Sciences, Hungarian Academy of Sciences, H-1121 Budapest, Hungary. 
\and LESIA, CNRS, Universit\'e Pierre et Marie Curie, Universit\'e Denis Diderot, Observatoire de Paris, 92195 Meudon cedex, France. 
\and NASA Ames Research Center, Moffett Field, CA 94035, USA. 
\and School of Physics and Astronomy, University of Birmingham, Edgebaston, Birmingham B13 2TT, UK. 
\and Department of Astronomy and Astrophysics, Villanova University, 800 East Lancaster Avenue, Villanova, PA 19085, USA. 
\and Institut d'Astronomie et d'Astrophysique, Univ. Libre de Bruxelles, CP226, Boulevard du Triomphe, 1050 Brussels, Belgium. 
\and Royal Observatory of Belgium, Ringlaan 3, 1180 Brussels, Belgium.
}

\abstract
{The unparalleled photometric data obtained by NASA's \textit{Kepler} space telescope led to an improved understanding of red giant stars and binary stars. Seismology allows us to constrain the properties of red giants. In addition to eclipsing binaries, eccentric non-eclipsing binaries, exhibiting ellipsoidal modulations, have been detected with \textit{Kepler}.} 
{We aim to study the properties of eccentric binary systems containing a red giant star and derive the parameters of the primary giant component.}
{We apply asteroseismic techniques to \new{determine} masses and radii of the primary component of \new{each} system. For a selected target, light and radial velocity curve modelling techniques are applied to extract the parameters of the system and its primary component. Stellar evolution and its effects on the evolution of the binary system are studied from theoretical models.}
{The paper presents the asteroseismic analysis of 18 pulsating red giants in eccentric binary systems, for which masses and radii were constrained. The orbital periods of these systems range from 20 to 440 days. The results of our ongoing radial velocity monitoring program with the \textsc{HERMES} spectrograph reveal an eccentricity range of $e$\,=\,0.2 to 0.76.
As a case study we present a detailed analysis of  KIC\,5006817, \new{whose} rich oscillation spectrum allows for a detailed seismic analysis.
From seismology we constrain the rotational period of the envelope to be at least 165\,d, which is roughly twice the orbital period. The stellar core rotates 13 times faster than the surface. From the spectrum and radial velocities we expect that the Doppler beaming signal should have a maximum amplitude of 300\,ppm in the light curve. \new{Fixing the mass and radius to the asteroseismically determined values, from our binary modelling we find a value of the gravity darkening exponent that is significantly larger than expected. Through binary modelling, we determine the mass of the secondary component to be 0.29$\pm$0.03\,$M$\sun.}}
{For KIC\,5006817 we exclude pseudo-synchronous rotation of the red giant with the orbit. The comparison of the results from seismology and modelling of the light curve shows a possible alignment of the \new{rotational and orbital axis} at the 2$\sigma$ level. Red giant eccentric systems could be progenitors of cataclysmic variables and hot subdwarf B stars. }


\maketitle

\section{Introduction}
\begin{figure*}[t!]
\includegraphics[width=\hsize]{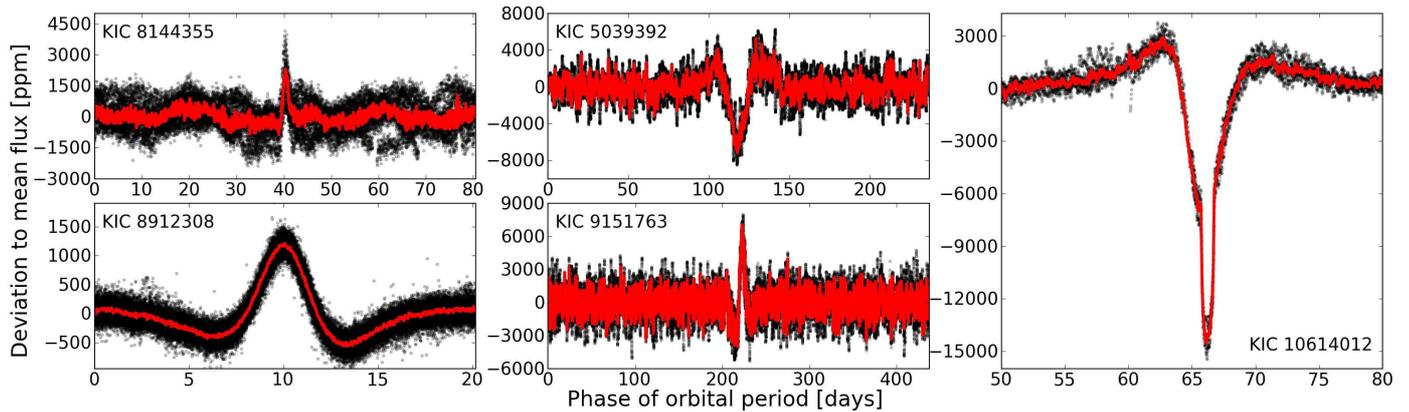}
\caption{\label{fig:rghbEnsemble} Five examples of \textit{Kepler} lightcurves of red giant heartbeat stars from our sample. 
Black dots and red line show the corrected and rebinned (30 min) lightcurves, respectively.
\object{KIC\,8144355} is the star with the highest eccentricity, \object{KIC\,5039392} is the most luminous star, and \object{KIC\,9151763} is the star with the longest orbital period. 
The low\new{-luminosity} RGB star \object{KIC\,8912308} has the shortest orbital period in our sample. The right panel shows a zoom on the  partial eclipse during the heartbeat event of \object{KIC\,10614012}.  }
\end{figure*}

The NASA \textit{Kepler} space telescope \citep{borucki2010} has been delivering unprecedented photometric data for more than 150\,000 stars. These nearly continuous observations that now cover more than 1000\,d have allowed major advances in our understanding of stellar structure of single stars and of multiple-star systems. The asteroseismic investigation of red giant stars has become one of \textit{Kepler's} success stories.

Red giants are evolved stars that have ceased hydrogen burning in the core and left the main sequence. This late phase of stellar evolution splits into several episodes, which are associated with subsequent modifications of the stellar structure. As the helium core contracts, the energy generation rate in the hydrogen-burning shell that surrounds the core increases the luminosity causing an increase in the stellar radius. The core's mass and density keep growing until the core is hot enough to ignite helium, as the helium settles from the burning hydrogen shell. For stars with birth masses below approximately 2.3\,$M_\odot$, the helium core fusion starts with a helium flash in a series of off-centre thermal sub-flashes \citep[e.g.][]{bildsten2012} that result in a thermal runaway. At this point the core expands by two orders of magnitude, reducing the temperature of the overlying hydrogen-burning shell. The luminosity of the star drops, the radius contracts and the red giant settles onto the horizontal branch, where metal rich stars are concentrated in the red clump.

A large fraction (in radius) of the red giant's envelope is convective and oscillations are excited stochastically in a part of that envelope. These solar-like oscillations correspond to pressure modes (p\,modes) and their frequencies follow a characteristic comb-like pattern \citep{tassoul1980}. The mode amplitudes range from a few tens to a few hundred parts per million (ppm) in observed flux, or of the order of 10\,m\,s$^{-1}$ or less in radial velocity \citep{frandsen2002, deridder2006}, which is barely detectable from ground-based observations. In the era before high-precision photometric space missions, solar-like oscillations of red giants were only confirmed in a handful of bright stars from extensive single- and multisite campaigns of high-precision spectroscopy.  Examples are: \ $\xi\,$Hya \citep{frandsen2002}, $\varepsilon\,$Oph \citep{deridder2006}, and $\eta\,$Ser \citep{hekker2006}. Photometric measurements from space have substantially increased the number of red giants with detected oscillation signals. The detection of nonradial modes in a multitude of red giants observed with the \textsc{CoRoT} satellite was a major milestone to an improved understanding of red giants, as it allowed for more sophisticated asteroseismic analyses \citep{deRidder2009}.

Pressure modes have their largest amplitude in the convective envelope while gravity modes (g\,modes) have their largest amplitude in the deep interior. At some stages of stellar evolution, p\,modes and g\,modes can couple and become a mixed mode. The firm identification of such mixed modes in \textit{Kepler} and \textsc{CoRoT} data by \cite{beck2011}, \cite{bedding2011} and \cite{mosser2011a} extended the sensitivity of the seismic analysis towards the core regions of red giant stars \citep[][and references therein]{dupret2009}. The analysis of mixed modes allows us to determine the evolutionary state \citep{bedding2011, mosser2011a} and constrain the core rotation of red giants \citep{beck2012, deheuvels2012}.

Seismology of red giants is largely built on scaling relations for pressure modes that were already described by \cite{kjeldsen1995}. These relations have been applied in numerous papers since then \citep[e.g.][]{kallinger2010, huber2011,corsaro2013}. Results from scaling relations were recently confronted with results of independent methods. \cite{huber2012ApJ} found a good agreement within the observational uncertainties for evolved stars. 
\new{From comparing results for the eclipsing binary \object{KIC\,8410637} from binary modelling and seismology, \cite{frandsen2013} found an excellent agreement for the surface gravity. But, the masses determined by the two methods deviated slightly.}
Additionally, a large set of red giants in eclipsing binaries was recently reported by \cite{Gaulme2013} which will allow for new sample  studies.

The \textit{Kepler} mission has also recently discovered the existence of a new class of \new{eccentric ellipsoidal} binary stars for which the \new{binary characteristics can be determined over the complete range of inclinations and as such are not} limited to the narrow range of eclipsing systems. These so-called \textit{Heartbeat} stars \citep{welsh2011, thompson2012} are defined as eccentric, detached binary systems that undergo strong gravitational distortions and heating during periastron passage, which are clearly depicted in their lightcurves (e.g., Fig.\,1). \cite{kumar1995} developed a theory of such objects and demonstrated how the morphology of the light curve is defined by the eccentricity, argument of periastron and inclination. Furthermore, the amplitude of the periastron variation is determined by the periastron separation, masses and structure of the stellar components. Consequently, these parameters can in principle be gauged through consideration of the light and radial velocity curves. The first such object discovered to confirm the theory was KOI\,54 (HD\,187091, \citealt{welsh2011}). Subsequently, more tidally interacting eccentric binary systems have been discovered in the \textit{Kepler} field as well as in the Magellanic clouds from OGLE observations by \cite{thompson2012}, \cite{Hambleton2013}
 and \citet{nicholls12}, respectively.
Only observations from \textit{Kepler} provide the temporal resolution to allow for a dedicated seismic analysis for these stars. Table \ref{tab:asteroseismicValues} lists several newly-found heartbeat stars in the \textit{Kepler} field of view with a component exhibiting solar-like oscillations. 

\begin{figure}[t!]
\includegraphics[width=\hsize]{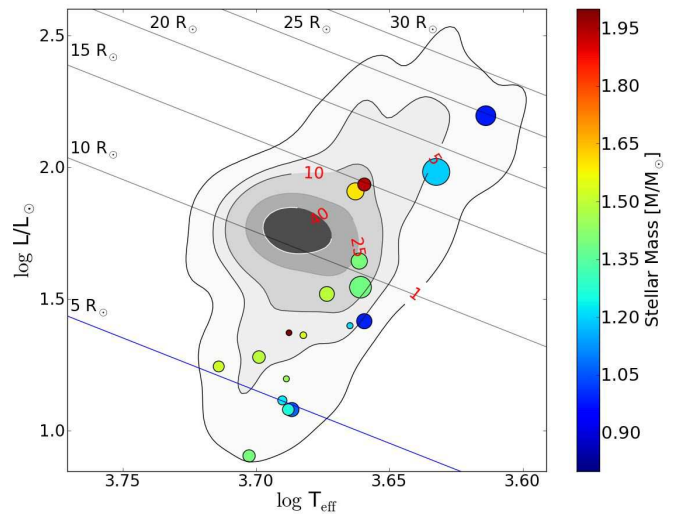}
\caption{Position of the 18 red giant heartbeat stars from Tables\,\ref{tab:asteroseismicValues} and \ref{tab:shortCadenceGiants} in the HR~Diagram, where the colour shows the mass of the red giant, derived from seismology. The size of the dots represents the orbital period, ranging between 20 to 438\,d. The contour surfaces reflect the density distribution of 1000 pulsating red giants. The darkest areas mark the position of the densely populated red clump. Numbers in red indicate the star count per bin, for which the contour surfaces have been drawn. Lines of equal radii in the HR~Diagram have been drawn for selected stellar radii between 5 and 30\,$R$\sun.
\label{fig:rghbHRD} }
\end{figure}

The seismic study of a sample of 18 eccentric systems (Fig.\,\ref{fig:rghbHRD}) is presented in Section 3. For a detailed seismic study, a large number of \new{oscillation} modes is needed to allow for \new{an optimal} comparison with theoretical models. The star with the richest power density spectrum is KIC\,5006817. Section 4, 5 and 6 describe the seismic, spectroscopic and binary analyses of this system. The results of these different analyses are compared in Section 7. In Section\,\ref{sec:sdBDiscussion} we reflect upon the possibility that heartbeat stars are potential subdwarf B (sdB) and cataclysmic variables (CV) progenitors.

\section{Observations 
\label{sec:Observations}}

\begin{table*}[t!]
\caption{\label{tab:asteroseismicValues}
\new{Seismic and fundamental parameters} for 14 oscillating red giant heartbeat stars, ordered by descending orbital period.}
\centering
\tabcolsep=3.5pt
\begin{tabular}{c|rrrrrrrrrrrrrrrrrrr}
\hline \hline
\multicolumn{1}{c}{Star}  & \multicolumn{1}{c}{$\nu_{\rm max}$}  & 
\multicolumn{1}{c}{$\Delta\nu$}  & \multicolumn{1}{c}{$\Delta\Pi_1$}  & 
\multicolumn{1}{c}{$\delta f_{\rm max}$} &  \multicolumn{1}{c}{Evol.}  & 
\multicolumn{1}{c}{R}  & 
\multicolumn{1}{c}{M}  & 
\multicolumn{1}{c}{$\log g$} & 
\multicolumn{1}{c}{L}  & 
\multicolumn{1}{c}{T$_{\rm eff}$} & 
\multicolumn{1}{c}{$P_{\rm orbit}$} & 
\multicolumn{1}{c}{A}  & 
\multicolumn{1}{c}{$| \Delta $RV$|$} 	 \\
\multicolumn{1}{c}{KIC}  & 
\multicolumn{1}{c}{ [$\mu$Hz]} & 
\multicolumn{1}{c}{ [$\mu$Hz]} & 
\multicolumn{1}{c}{ [sec]} & 
\multicolumn{1}{c}{ [nHz]} & 
\multicolumn{1}{c}{Phase}  & 
\multicolumn{1}{c}{ [$R$\sun]} & 
\multicolumn{1}{c}{ [$M$\sun]} & 
\multicolumn{1}{c}{ [dex]} & 
\multicolumn{1}{c}{ [L\sun~]} & 
\multicolumn{1}{c}{ [K]} & 
\multicolumn{1}{c}{ [d]} & 
\multicolumn{1}{c}{ [ppt]} & 
\multicolumn{1}{c}{ [km\,s$^{-1}$]}
\\\hline

\hline
{9151763}	 & 13.8$\pm$0.2 & 1.98$\pm$0.01 	& $-$ & $-$ &  RGB? & 17.6$\pm$0.4 & 1.19$\pm$0.08 & 2.01 & 96$\pm$16 & 4290 & 437.5 & +7.1 & 32.2\\
{7431665}	 &54.0$\pm$0.7 & 5.46$\pm$0.02& $\sim$67 & $-$ &  RGB & 9.4$\pm$0.1 & 1.39$\pm$0.05 & 2.62 & 35$\pm$2 & 4580 & 281.4&-3.0 &[37.8]\\
{5039392} & 6.2$\pm$0.1 & 1.13$\pm$0.01 & $-$ & $-$ & RGB &  24.0$\pm$0.7 & 0.98$\pm$0.07 & 1.67 & 157$\pm$24 & 4110& 236.7 & -6.0 &42.3\\
{9540226}$^\star$ & 27.4$\pm$0.4 & 3.18$\pm$0.01 & $-$ & $-$ & RGB & 14.1$\pm$0.3 & 1.6$\pm$0.1 & 2.37 & 81$\pm$13 & 4600 & 175.4 &  -7 & 45.3\\
{8210370}	 &44.1$\pm$0.8 & 4.69$\pm$0.02	& $-$ & $-$ &  RGB? & 10.5$\pm$0.2 & 1.40$\pm$0.08 & 2.54 & 44$\pm$4  & 4585 & 153.5 & -5.3 & 22.1\\
{11044668}&50.2$\pm$0.2 & 5.65$\pm$0.01	& $\sim$60 & 83(?) &  RGB & 8.18$\pm$0.09 & 0.99$\pm$0.03 & 2.59 & 26$\pm$3 & 4565 & 139.5 & -3.8& [43.0]\\
{10614012$^\star$}&70.2$\pm$0.9	& 6.54$\pm$0.02 	& $-$ & $-$&  RGB & 8.6$\pm$0.2& 1.49$\pm$0.08 & 2.74 & 33$\pm$4 & 4715 & 132.1 & -4.7 &  49.3\\
{9163796}	 &153.2$\pm$0.7& 13.53$\pm$0.04	& $-$ & $-$&  RGB & 4.46$\pm$0.03 & 0.89$\pm$0.01 & 3.09 &  12$\pm$1  & 4820 & 121.3 &$\pm$0.5& 70.1 \\
{2444348} & 30.5$\pm$0.3& 3.26$\pm$0.01  	& $-$ & $-$ &  RGB & 14.9$\pm$0.3 &1.94$\pm$0.11 & 2.38 & 86$\pm$14 & 4565 & 103.5 & -1.7 & 7.7\\
\textbf{5006817}& 145.9$\pm$0.5 & 11.64$\pm$0.01 & 78 & 450 &  RGB & 5.84$\pm$0.09 & 1.49$\pm$0.06 & 3.08 & 19$\pm$3 & 5000 & 94.8 & -1.7& 23.5\\
8803882 & 347$\pm$3& 22.6$\pm$0.4& $-$ & 500(?) & RGB   & 3.68$\pm$0.1  & 1.4$\pm$0.1 & 3.45& 8$\pm$1& 5043 & 89.7 & +0.5 & [1.9]\\
{8144355}	 & 179$\pm$2& 13.95$\pm$0.04	& $\sim$78 & 210(?)&  RGB & 4.90$\pm$0.09 & 1.26$\pm$0.08 &3.16 & 12$\pm$2  & 4875& 80.6 & +2.1 & 18.9\\
{9408183} & 164.8$\pm$0.2& 13.29$\pm$0.02 & $\sim$93 & 450 &  RGB & 5.02$\pm$0.07 & 1.23$\pm$0.05 & 3.12 & 13$\pm$1  & 4900 & 49.7 & +1.5& 64.4\\
{2720096} & 110.1$\pm$0.7& 9.17$\pm$0.01 & $-$ & $-$ & RGB   & 6.98$\pm$0.08 & 1.54$\pm$0.06 & 2.95 & 23$\pm$2 & 4812 & 26.7 & +1.0& 4.0 \\
{8095275} & 69.3$\pm$0.3  & 6.81$\pm$0.01 & $-$ & $-$ & RGB & 7.78$\pm$0.08 & 1.21$\pm$0.05 & 2.74 & 25$\pm$3 & 4622 & 23.0 & -6.0 & 20.6 \\
\hline
\end{tabular}
\tablefoot{The star's identifier in the \textit{Kepler} Input Catalogue (KIC) is given. Eclipsing systems are marked  with an asterisk. The columns $\nu_{\rm max}$ and $\Delta\nu$ report the frequency of the oscillation power excess and the large \new{freq}uency separation between radial modes for a given star. $\Delta\Pi_1$ quantifies the true period spacing of dipole modes. The maximum value of the detected rotational splitting $\delta f$ is listed. The evolutionary phase RGB describes H-shell burning red giant. \new{Ambiguous values are marked with '?'.}
The columns $R$,\,$M$,\,$L$, and $\log g$ report the stellar radius, mass, luminosity, effective temperature and surface gravity from scaling relations, respectively. 
\new{$T_{\rm eff}$ was adopted from the KIC.}
The uncertainties of $\log g$ are on the order of 0.01\,dex and for the temperature typically smaller than 150\,K.
$P_{\rm orbit}$ gives the orbital period from photometry.  
The column $A$ lists the maximum amplitude of the heartbeat in a rebinned phase diagram. 
\new{The error estimate for $P_{\rm orbit}$  and $A$ from the PDM is not reliable due to the remaining contamination of the solar-like oscillations and therefore not given.}
$| \Delta $RV$|$ reports the maximum difference in radial velocity. \new{Squared brackets mark systems for which the orbital parameters could not yet be determined from radial velocities.} }
\\ \vspace{7mm}
\caption{\new{Low-luminosity} red giants with $\nu_{\rm max}$ higher than 283\,$\mu$Hz from long cadence data.
 \label{tab:shortCadenceGiants}}
 \centering

\begin{tabular}{c|rrrrrrrrrrr}
\hline\hline
\multicolumn{1}{c}{Star}  &
\multicolumn{1}{c}{$f_{\rm max}$} &
\multicolumn{1}{c}{$(\nu_{\rm max})$}	& 
\multicolumn{1}{c}{$\Delta\nu$} & 
\multicolumn{1}{c}{R} &			
\multicolumn{1}{c}{M} &		
\multicolumn{1}{c}{$\log g$} &
\multicolumn{1}{c}{L } &
\multicolumn{1}{c}{T$_{\rm eff}$} &
\multicolumn{1}{c}{P$_{\rm orbit}$ } &
\multicolumn{1}{c}{A } &
\multicolumn{1}{c}{$| \Delta$RV$|$ } \\

\multicolumn{1}{c}{KIC}&
\multicolumn{1}{c}{[$\mu$Hz]}&
\multicolumn{1}{c}{[$\mu$Hz]} &
\multicolumn{1}{c}{[$\mu$Hz]} &
\multicolumn{1}{c}{[$R$\sun]}&
\multicolumn{1}{c}{[$M$\sun]}&
\multicolumn{1}{c}{[dex]}&
\multicolumn{1}{c}{[L\sun]}&
\multicolumn{1}{c}{[K]}&
\multicolumn{1}{c}{[d]}& 
\multicolumn{1}{c}{[ppt]}&
\multicolumn{1}{c}{[km\,s$^{-1}$]}  \\ \hline

7799540	& 220$\pm$5 	& (347.2)			&24.0& 3.64	& 1.52	& 3.50	&17.5
& 5177	& 71.8 	& +0.5 	& [31.8]  \\
2697935$^{\star}$	& 161$\pm$3 	& (405.6)			&$\sim$28& 3.26	& 1.45	& 3.574	& 15.7 
& 4883 	& 21.5	&$\pm$1.3 & 52.1 \\ 
8912308	& 217$\pm$9 	& (350.2)			&22.7& 4.20	& 2.02	& 3.50	& 23.5
& 4872	& 20.2	&+1.2	&61.4 \\\hline
\end{tabular}
\tablefoot{ The definition of columns $KIC$,  $\Delta\nu$, $M$, $R$, $\log g$, $L$, $P_{\rm orbit}$, $A$ and $\Delta|$RV$|$ is the same as in Table\,\ref{tab:asteroseismicValues}. $f_{\rm max}$ and $(\nu_{\rm max})$ indicate the frequency of the maximum oscillation power, reflected at the Nyquist frequency and reconstructed power excess, respectively. $T_{\rm eff}$ was adopted from the KIC-parameters. The uncertainties of $M$, $R$, $\log g$ and $L$ are better than 2, 5, 1, and 1\new{5} percent, respectively.}
\\ \vspace{7mm}
\tabcolsep=4pt
\centering
\caption{\label{tab:sampleOrbits} 
Orbital parameters for systems for which  periastron has been monitored with the \textsc{HERMES} spectrograph. 
}
\begin{tabular}{c|rrrrrrrl}
\hline\hline
\multicolumn{1}{c}{Star} &
\multicolumn{1}{c}{$n_{\rm RV}$}  &
\multicolumn{1}{c}{$P_{\rm orbit}$}  &
\multicolumn{1}{c}{$e$}  &
\multicolumn{1}{c}{$\Omega$}  &
\multicolumn{1}{c}{$K$} &
\multicolumn{1}{c}{$\gamma$} &
\multicolumn{1}{c}{T$_0$} & 
\multicolumn{1}{l}{Eclipse} \\

\multicolumn{1}{c}{KIC} &
\multicolumn{1}{c}{}  &
\multicolumn{1}{c}{[d]}  &
\multicolumn{1}{c}{}  &
\multicolumn{1}{c}{[rad]}  &
\multicolumn{1}{c}{[km/s]} &
\multicolumn{1}{c}{[km/s]} &
\multicolumn{1}{c}{[HJD]} &
\multicolumn{1}{l}{duration}\\

\hline
9151763	& 24	& 437.51$\pm$0.03	& 0.73$\pm$0.01	& 3.03$\pm$0.01	& 16.20$\pm$0.04	& -92.89$\pm$0.03 	& 2455949.64$\pm$0.06	\\
5039392		& 13 & 236.70$\pm$0.02 	& 0.44$\pm$0.01	& 4.96$\pm$0.01	& 22.6$\pm$0.2		& -14.96$\pm$0.05 	& 2454874.2$\pm$0.27	\\
9540226$^\star$		& 31	& 175.43$\pm$0.01	& 0.39$\pm$0.01	& 0.07$\pm$0.01	& 23.32$\pm$0.04	& -12.37$\pm$0.02 	& 2456425.89$\pm$0.09	& P: 4\,d; S: 3\,d \\
8210370 & 16 & 153.50$\pm$0.01 & 0.70$\pm$0.01 & 1.17$\pm$0.01 & 12.96$\pm$0.36 & -0.76$\pm$0.08 & 2454937.35$\pm$0.01 \\
10614012$^\star$		& 22	& 132.13$\pm$0.01	& 0.71$\pm$0.01 	& 1.23$\pm$0.01 	& 24.68$\pm$0.03 	& -0.92$\pm$0.02  	&  2454990.48$\pm$0.01 & 1\,d 	\\
9163796 		& 17	& 121.30$\pm$0.01 	& 0.69$\pm$0.01	& 0.00$\pm$0.01 	& 35.64$\pm$0.01 	& -11.05$\pm$0.01	& 2456409.60$\pm$0.01 \\
2444348		& 17 & 103.50$\pm$0.01	& 0.48$\pm$0.01	& 4.30$\pm$0.01 	& 4.76$\pm$0.02	& 14.47$\pm$0.01 	& 2454947.74$\pm$0.09	\\
{\bf 5006817}	& 70 & 94.812$\pm$0.002& 0.7069$\pm$0.0002& 4.0220$\pm$0.0005	& 11.709$\pm$0.005	& -14.021$\pm$0.002		& 2456155.924$\pm$0.002	\\
8144355 		& 19	& 80.55$\pm$0.01	& 0.76$\pm$0.01 	& 2.79$\pm$0.01 	& 9.44$\pm$0.04	& 0.02$\pm$0.03	& 2455914.43$\pm$0.02	\\
9408183 & 7 & 49.70$\pm$0.01 & 0.42$\pm$0.01 & 0.17$\pm$0.01 & 37.17$\pm$0.04 & -14.37$\pm$0.01 & 2454989.80$\pm$0.01 \\
2720096		& 13 & 26.70$\pm$0.01 	& 0.49$\pm$0.01 	& 6.11$\pm$0.01 	& 2.28$\pm$0.03	& 9.92$\pm$0.01 	& 2454990.67$\pm$0.05	\\
8095275		& 25	& 23.00$\pm$0.01	& 0.32$\pm$0.01	& 2.19$\pm$0.01 	& 10.50$\pm$0.06	& -8.58$\pm$0.03 	& 2454971.02$\pm$0.03	\\
2697935${^\star}$		& 27 & 21.50$\pm$0.01 	& 0.41$\pm$0.02 	& 2.33$\pm$0.06 	& 26.5$\pm$0.6		& -74.4$\pm$0.4 	& 2454990.9$\pm$0.1 & 0.1\,d	\\
8912308		& 28 & 20.17$\pm$0.01 	& 0.23$\pm$0.01 	& 3.34$\pm$0.01 	& 30.78$\pm$0.02	& -52.69$\pm$0.01 	& 2454994.06$\pm$0.01	\\
\hline
\end{tabular}
\tablefoot{Number of radial velocity measurements, $n_{\rm RV}$. The orbital period $P_{\rm orbit}$, the eccentricity $e$, the argument of periastron $\Omega$, the radial velocity amplitude $K$, the velocity of the system $\gamma$, and the zero point $T_0$. The duration of the eclipse is given.  P and S indicate the primary and secondary eclipse, respectively.}
\end{table*}

The \textit{Kepler} datasets used in this study cover a time base up to 1300\,d (Quarters Q0--Q14) in the long cadence observing mode of \textit{Kepler}. In this mode, integrations of 6.04\,s are taken every 6.54\,s and 270 such integrations are co-added for 
transmission to Earth to give an integration time of 29.43\,min, leading a Nyquist frequency of 283.4\,$\mu$Hz. For one target a month of short cadence \new{data} was available. In this observing mode, the individual integrations are stacked to exposures of 58.8\,s, which leads to a Nyquist frequency of 8495\,$\mu$Hz (Fig.\,\ref{fig:superNyquist}).
To produce lightcurves which are robust against long period instrumental drifts, we extracted the photometric flux from the pixel data following the methods described by \cite{bloemen2013}. The lightcurves were corrected following \cite{garcia2011}. The target pixel data were also used to inspect whether the light curve was contaminated by neighbouring field stars. In all cases a significant contamination is unlikely.

Red giant heartbeat candidates were selected \new{by} inspecting the lightcurves. Also the subsequent versions of the \textit{Kepler} eclipsing binary catalogue \citep{Prsa2011,slawson2011,matijevic2012} were searched for candidates among the stars classified as red giants.

The lightcurves of red giant stars are dominated by the low frequency signals of the granulation background and of the solar-like oscillations (Fig.\,\ref{fig:panoramixFullPDS}), which hamper a determination of the precise value of the orbital period from the reoccurring flux modulation at periastron (hereafter referred to as \textit{heartbeat event}). Therefore, the lightcurves were smoothed with a boxcar with a width of a few days and analyzed using the phase dispersion minimization technique \citep{stellingwerf}. By testing for strictly periodic recurrences, confusion with other variability sources, {such as stellar activity or instrumental artifacts is excluded.}

For an independent confirmation of the binary nature of the discovered heartbeat systems, we searched for radial velocity variations from spectra obtained with the \textsc{HERMES} spectrograph \citep{raskin2011}, mounted on the 1.2-m \textsc{Mercator} telescope at La Palma, Canary Islands, Spain. This highly efficient \'echelle spectrograph has a resolving power of $R=86\,000$. The raw spectra were reduced with the instrument-specific pipeline. The radial velocities were derived through weighted cross-correlation of the wavelength range between 478 and 653\,nm of each spectrum with an Arcturus template \citep{raskin2011}. For our prime target, KIC\,5006817, two orbital cycles were monitored in 2012 \new{as well as in 2013}. Radial velocity monitoring for the other stars listed in Table\,\ref{tab:asteroseismicValues} is ongoing. Tables \ref{tab:asteroseismicValues} to \ref{tab:sampleOrbits} report the first orbital results for stars in our sample.

\section{Eccentric red giant systems in the \textit{Kepler} data 
\label{sec:rghbStars}}

In total, we found 18 red giant stars in eccentric binary systems that show the characteristic gravitational distortion of heartbeat stars.  These stars were found in a sample of \textit{Kepler} red giants that encompass about 16\,000 stars. All stars show the clear signature of solar-like oscillations (Tables\,\ref{tab:asteroseismicValues} and \ref{tab:shortCadenceGiants}). 
Fig.\,\ref{fig:rghbEnsemble} shows phase diagrams of five selected stars. 

The global seismic analysis of red giant stars enables us to accurately constrain the fundamental parameters of the main component of the binary system. The characteristic comb-like structure of solar-like oscillations originates from p\,modes and therefore depends on the sound speed in the acoustic cavity. The main characteristics of the power excess (central frequency of the oscillation power excess, $\nu_{\rm max}$, and the large frequency separation between consecutive radial modes, $\Delta\nu$) scale well with the mass and radius, and to a lesser extent with effective temperature \new{\citep[e.g.,][]{kjeldsen1995,Mosser2013}}. Following the approach of \cite{kallinger2010} we estimated the radius, mass, luminosity, $\log g$ and effective temperature of our sample stars (Tables\,\ref{tab:asteroseismicValues} and \ref{tab:shortCadenceGiants}) and place the red giant components of the binaries in the Hertzsprung-Russell (HR)~diagram shown in Fig.\,\ref{fig:rghbHRD}. All stars except two have masses between 1 and 1.5\,$M$\sun. The majority of the \textit{Kepler} red giants are located in this mass range.

\subsection{The evolutionary status of red giant heartbeat stars}
Seismic information can reveal the evolutionary status of the star. Stars with a large frequency separation $\Delta\nu \gtrsim $\,9\,$\mu$Hz (i.e. R$_{\rm RG}\, \lesssim\,7\,R$\sun) are located well below the red clump (Fig.\,\ref{fig:rghbHRD}) and therefore can only be in the hydrogen shell burning phase. A value of $\Delta \nu \, \lesssim $ 3\,$\mu$Hz (i.e. R$_{\rm RG} \gtrsim\,13\,R$\sun) occurs for stars located above the red clump. Such stars are likely to be H-shell burning stars, high up on the RGB, although in principle they also could be stars on the low asymptotic giant branch (AGB). Stars with frequency separation 3 $\, \lesssim $\,$\Delta \nu$\,$\lesssim $ 9\,$\mu$Hz can also be in the helium core burning phase \citep{bedding2011,mosser2012b}. Therefore, a further criterion to determine the evolutionary state of stars with radii larger than 7\,$R$\sun~is needed.

The period separation of mixed dipole modes is a powerful diagnostic to discriminate between hydrogen shell and helium core burning stars \citep[c.f.][]{bedding2011, mosser2011b}. 
 The true period spacing $\Delta\Pi_1$ describes the constant period spacing of pure dipole g modes, which cannot be observed directly. However, one can determine the value of $\Delta\Pi_1$ by fitting a theoretical mixed mode pattern to the actually observed modes in the power spectrum \citep{mosser2012c}. Stars which exhibit a value of $\Delta\Pi_1\,\lesssim\,$100\,s are burning hydrogen in a shell around the inert helium core, while stars with a larger $\Delta\Pi_1$ value belong to stars on the AGB or RC. The estimated values of the true period spacing $\Delta\Pi_1$ for the pulsators are given in Table\,\ref{tab:asteroseismicValues}.

\begin{figure}[t!]
\centering
\includegraphics[width=\hsize]{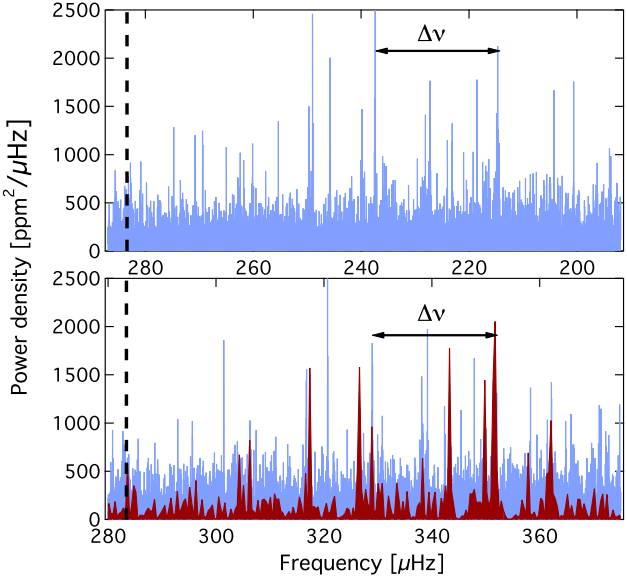}
\caption{Power density spectra of \object{KIC\,8803882} from 1250\,d of long and 30\,d of short cadence data is shown in blue and red, respectively. The formal Nyquist frequency (dashed line) separates the reflected oscillation power (top panel, sub Nyquist frequency range) and the original power excess (bottom panel, super Nyquist frequency range). \label{fig:superNyquist} }
\end{figure}

Several stars do not or barely show dipole mixed modes and it is impossible to recover their mixed mode pattern. 
For these stars we did not determine the evolutionary state from the period spacing but used the phase shift ($\epsilon_c$) of the central radial mode \new{as an indicator}. \cite{kallinger2012} have shown that in a diagram of $\epsilon_c$ versus the large separation $\Delta \nu$, the stars fall into 
groups which can be identified as H-shell burning, He-core burning and AGB stars. \new{For most stars, the identification of the evolutionary stage from $\epsilon_c$ and the observed period spacing \new{are in agreement}.}
Therefore we are also able to \new{estimate} the evolutionary state of stars which did not show a clear forest of dipole modes.

For nearly all heartbeat red giants we could constrain them to be in the state of H-shell burning. The remaining stars are also likely to be in the same evolutionary state but we cannot rule out more evolved phases. 
Among our sample there are 5 stars in a $\Delta\nu$ range where the stars can either be an RGB or RC star. Statistically there is a chance of about 70 percent that a given red giant in this range burns He in its core \citep{kallinger2012}. Even though the number statistics are still poor we find clear preference for RGB primaries in heartbeat stars. In Section\,\ref{sec:sdBDiscussion}, we discuss that this could be a result of binary evolution.

\subsection{Giants with power excess above the formal Nyquist frequency}
In the power density spectrum of KIC\,8803882, we find a reverse combination of $l$=2 and 0 modes with respect to the known comb-like structure between 200\,$\mu$Hz and the LC-Nyquist frequency of 283\,$\mu$Hz with an apparent large separation of $\Delta\nu=\,22.7\,\mu$Hz (Fig.\,\ref{fig:superNyquist}). Stars which show similar structures in their power spectra are KIC\,8912308, KIC\,2697935 and KIC\,7799540 (cf. Table\,\ref{tab:shortCadenceGiants}).

As the dataset of KIC\,8803882 contained a month of short cadence observations (Q14.1), we could compare the analysis of the super Nyquist frequency range from long cadence data with the same, well resolved frequency range from short cadence data. This comparison, depicted in Fig.\,\ref{fig:superNyquist}, allows us to explore the frequency range above the formal Nyquist frequency, $\nu_{\rm Nyquist}\,=\,283\,\mu$Hz for long cadence data \citep{Murphy2013}. We determined the large separation $\Delta\nu$ through manual peakbagging of the long cadence data to be 22.7\,$\mu$Hz, which is in perfect agreement with the large separation of 22.65\,$\mu$Hz from short cadence data. The standard approach to determine the frequency of the power excess \citep[e.g.][]{kallinger2010} uses a simultaneous fit of the Gaussian envelope, multiple power laws for the background and a white noise component. In the super Nyquist frequency domain, we find an artificial background with increasing power density towards higher frequencies as the signal of the low frequency domain is mirrored. Therefore, we determined the position of $\nu_{\rm max}$ by fitting a broad Gaussian to the reflected signal and calculated the true frequency of the power excess, 
\begin{equation}
\nu_{\rm max}^{\rm true}= 2 \cdot \nu_{\rm Nyquist} - \nu_{\rm max}^{\rm reflected}.
\end{equation}
\begin{figure}[t!]
\includegraphics[width=\hsize]{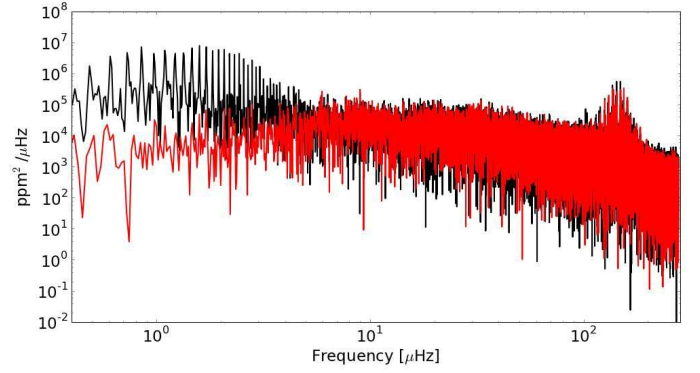}
\caption{Power density spectrum (PDS) of the light curve of KIC\,5006817. The black PDS was calculated from the original light curve. The red PDS originates from a high-passband filtered light curve, corrected for the mean flux variation during the periastron passage.  A zoom into the power excess is shown in 
Fig.\,\ref{fig:panoramixPDS}. 
\label{fig:panoramixFullPDS}}
\end{figure}

From the values obtained for KIC\,8803882, KIC\,8912308, KIC\,2697935, and KIC\,7799540 (cf. Tables\,\ref{tab:asteroseismicValues} and \ref{tab:shortCadenceGiants}) we conclude that they are low-luminosity red giants.

\begin{figure*}[t!]
\centering
\includegraphics[width=\hsize]{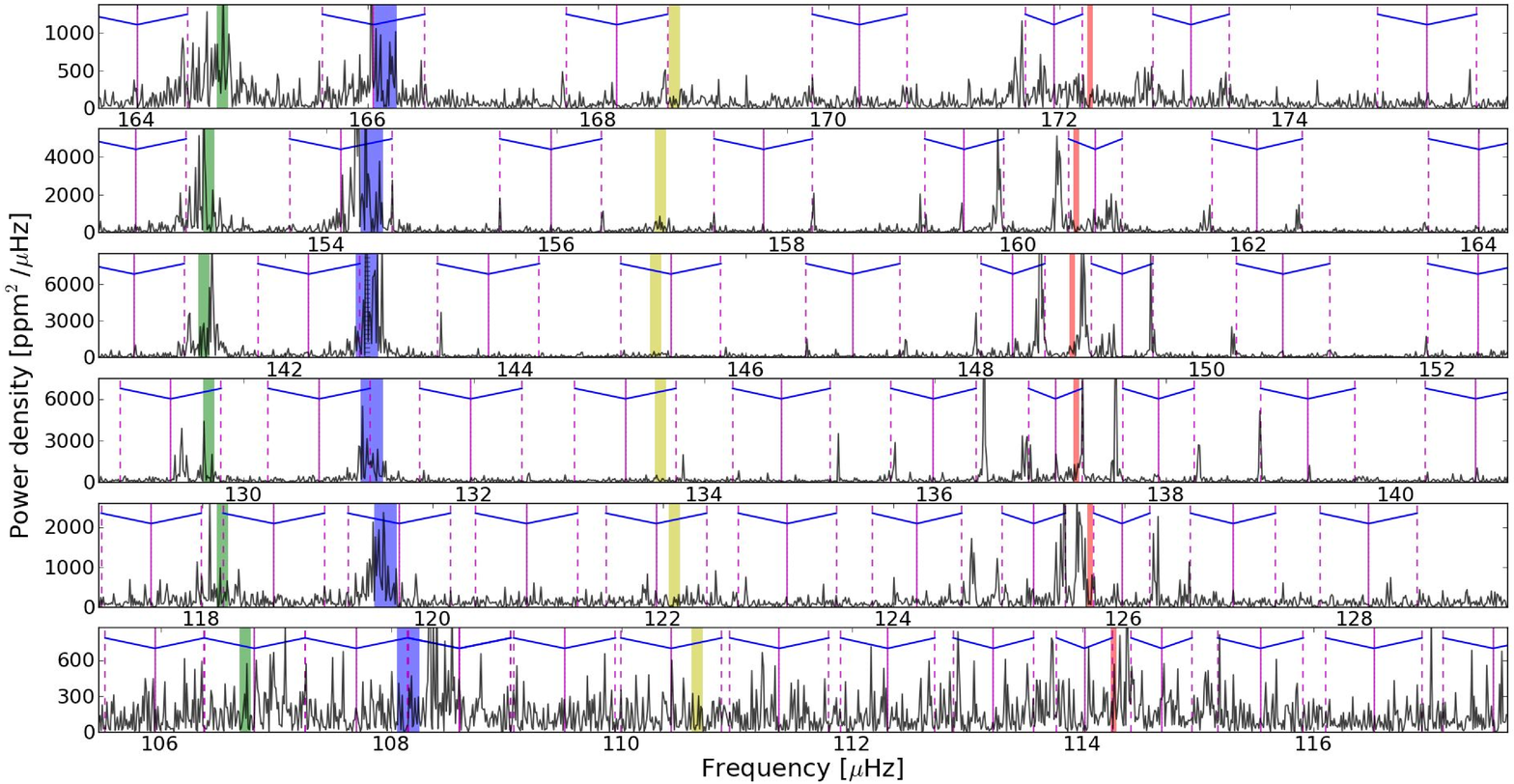}
\caption{ \label{fig:panoramixPDS} Power density spectrum (PDS) of KIC\,5006817 (Q0--Q13). Each panel \new{contains} one radial order. Mode identifications of the pure p\,modes come from the universal frequency pattern \citep{mosser2011b} for $\ell=0$, 1, 2 and 3 and are indicated with blue, red, green and yellow vertical bars, respectively. The effects of rotation are visible as the splitting of dipole modes, located in the centre of each panel. The observed PDS is overlaid with the theoretical frequencies of mixed dipole modes ($m$=0, solid thin lines) and the theoretical frequencies of the rotationally split components ($m$=$\pm$1, dashed thin lines) which have been calculated using methods described by \cite{mosser2012c, mosser2012b}. The components belonging to one rotationally split multiplet are indicated through $\rm V$-markers at the top of each panel. }
\end{figure*}

\section{The case study of KIC\,5006817
\label{sec:panoramixSeismology}}
While ensemble asteroseismology allows the easy characterization of large 
samples of red giant stars \citep[e.g.,][]{huber2011,hekker2011,kallinger2010,mosser2012c}, studies of individual stars in a close binary grant us \new{additional} insight into the structure of the primary as well as the interaction between the binary system components. As a proof-of-concept of what such objects have to offer compared to single pulsators, we organised a spectroscopic campaign on the heartbeat star with the richest oscillation pattern, KIC\,5006817.

The power density spectrum of the \textit{Kepler} observations of KIC\,5006817 is shown in Fig.\,\ref{fig:panoramixFullPDS}. It contains power excess centred around 146\,$\mu$Hz. Apart from the typical granulation signal and a series of low frequency peaks ($<$10\,$\mu$Hz) originating from the heartbeat event, no other significant frequencies are present (Fig.\,\ref{fig:panoramixFullPDS}). The oscillation power excess itself is typical for a red giant primary. The frequency range of the excited oscillation modes is shown in Fig.\,\ref{fig:panoramixPDS}. For a detailed asteroseismic analysis, only modes with a signal of at least 8 times the background level were extracted. The frequencies of the individual modes were extracted as the centroid of the power density in narrow predefined windows, which we checked for consistency by fitting Lorentzian profiles to a number of modes.

KIC\,5006817's seismic mass and radius estimates are $1.49 \pm 0.06\,M_\odot$ and $5.84 \pm 0.09\,R_\odot$, respectively (Table\,\ref{tab:asteroseismicValues}). The  true period spacing of 78\,s indicates that the star is in the phase of H-shell burning \citep{bedding2011,mosser2012b}.  In the HR~Diagram shown in Fig.\,\ref{fig:rghbHRD}, this star is located well below the red clump.

\subsection{Seismic information on the stellar rotation}
The power density spectra of many red giants contain clear signatures of the rotational splitting of nonradial modes, which enables us to learn more about the internal rotation of those stars. This effect arises as rotation breaks the degeneracy of nonradial modes by shifting the mode frequencies of the components with azimuthal order $m\neq0$ away from the central multiplet frequency ($m$=0):
\begin{equation}
f_{n,\ell,m} = f_{n,\ell,0} + \delta f_{n,\ell,m},
\end{equation}
with the rotational splitting $\delta f_{n,\ell,m}$ given by
\begin{equation}
\delta f_{n,\ell,m} = m\cdot\frac{\Omega}{2\pi}\cdot(1-C_{nl}),
\end{equation}
where $\Omega$ is the average rotation frequency in the cavity in which a given mode propagates and $C_{n\ell}$ is the Ledoux constant \citep{ledoux1951}. The values of the rotational splittings reported in this work are taken to be equal to the frequency separation between two consecutive multiplet components (i.e., $|\Delta m|$=1). If no central peak ($m$=0) is detected, we take half the value of the frequency difference between outer dipole components. We refer to such values as the normalized rotational splitting. In the power density spectrum of KIC\,5006817, we find rotationally split modes of the spherical degree $\ell = 1$. The multiplet structure in $l$ = 2 can originate from splitting or mixed modes. In principle, also $\ell=3$ modes should be split. 
However, we have no clear identification of them as in this star $l$=3 modes have amplitudes close to the significance limit.

\begin{figure}[t!]
\includegraphics[width=0.5\textwidth]{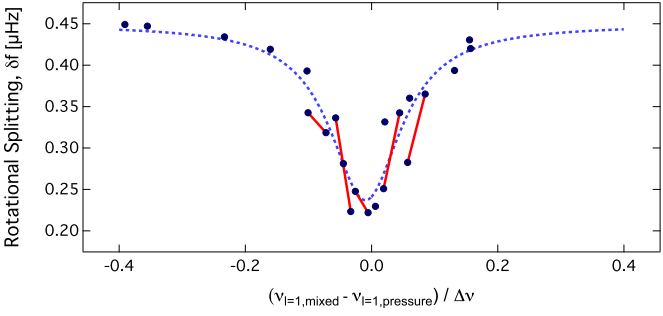}
\caption{\label{fig:splittingEchelle} 
Splitting-\'echelle diagram of rotationally affected dipole modes in KIC\,5006817. Measured rotational splittings are shown as dots. Solid lines connect two splittings originating from the same dipole mode triplet. The x-axis gives the position of a rotationally split dipole mode $\nu_{l=1, \rm mixed}$ with respect to the pure pressure dipole mode $\nu_{l=1, \rm pressure}$ and as fraction of the large frequency separation $\Delta\nu$ . The dashed line describes the modulation of the rotational splitting through a Lorentz-profile \citep{mosser2012c}. }
\end{figure}

\new{As g-dominated and p-dominated modes are sensitive to the rotation in the central and outer regions of the star, respectively, they can be used to probe the internal rotation gradient. \cite{beck2012} showed that larger splitting of g-dominated than of p-dominated dipole modes, as found in KIC\,5006817 (Figs.\,\ref{fig:splittingEchelle} and \ref{fig:modelPanoramix}) reveals that its interior, rotates multiple times faster than its envelope.}

\new{For some dipole multiplets the presence of a significant central peak ($m$\,=\,0) allowed us to measure the individual splittings for the $m$\,=\,$-1$ and $+1$ components, revealing asymmetries. Such pairs of splittings are connected with a solid line in Fig.\,\ref{fig:splittingEchelle}.
As rotation shifts the frequency of modes, each mode within a triplet has a slightly different oscillation cavity, which also modifies its mixed character in terms of p- and g-mode components. The asymmetries are mirrored around the pure p\, mode and follow the Lorentzian description (Fig.\,\ref{fig:splittingEchelle}).}

\new{The smallest measured rotational splitting of dipole modes is about 0.21\,$\mu$Hz and was measured in two asymmetric p-mode dominated mixed modes. This is about 2.25 times smaller than the largest splitting $\delta f_{\rm max}$ of 0.45\,$\mu$Hz found for g-mode dominated dipole modes. The largest splitting values ($\delta f_{\rm max}$) originating from the g-mode dominated dipole modes are consistent with those measured for large samples of single pulsating red giants \citep[e.g.][]{beck2012, mosser2012c}}

The extracted rotational splitting of dipole modes (Fig.\,\ref{fig:splittingEchelle}), shows the expected modulation as a function of the degree of mixed character \citep{mosser2012b, mosser2012c}. 
\newer{The mode identification from the universal red giant oscillation pattern for pressure modes \citep{mosser2012c}, and from asymptotic expansion for mixed modes and the rotational splittings \citep{mosser2012c, mosser2012b} is indicated with vertical bars in Fig.\,\ref{fig:panoramixPDS}.} 
\new{We note that a perfect fit is not needed to identify the modes.}

\subsection{Testing of the rotational profile from forward modelling}
Because of the conservation of angular momentum, one expects the core of a red giant to rotate significantly faster than its surface. Previous analyses \citep{beck2012, deheuvels2012, mosser2012c} only revealed the ratio of the core-to-surface rotation rate, rather than the shape of the rotational gradient in the transition region between the faster rotating core and the slower rotating envelope. \cite{goupil2013} propose a way to infer directly the ratio of the average envelope to core rotation rates from the observations. In this paper, we take a different approach by using forward modelling.

In principle, the number of rotationally split modes in KIC\,5006817 and their different degree of mixed character (and therefore different ``sensitivity'' to internal layers of the star) should allow to probe the rotation rate at different depths of the star, i.e., to resolve the rotational gradient to more than a ratio between the core and surface value. The radial structure of a red 
giant is dominated by a helium core and an extended convective envelope. To mimic this structure we considered models consisting of consecutive shells, which are each assumed to rotate rigidly, but with different angular velocities.

We computed such a representative 1.5\,$M_\odot$ model using the Yale Stellar Evolution Code \citep[YREC;][]{demarque2008, guenther1992}. The model was selected to approximately reproduce the observed  radial modes and true dipole period spacing of KIC\,5006817, where we note that no ``exact'' match is necessary as the mode eigenfunctions (and therefore the rotational kernels) of similar models are almost identical \citep[see also][]{deheuvels2012}. Our representative model was computed for near solar composition ($Z = 0.02, Y = 0.28$) assuming the solar mixture by \citet{grevesse1996} and a mixing length parameter ($\alpha_\mathrm{MLT} = 1.8$). The model has a radius, effective temperature, age, and inert He core mass fraction of about 5.8\,$R_\odot$, 4855\,K, 2.8\,Gyr, and 0.14, respectively. More details about the input physics of the model are given in \citet{kallinger2010b}.

\begin{figure*}[t!]
\centering
\includegraphics[width=\textwidth]{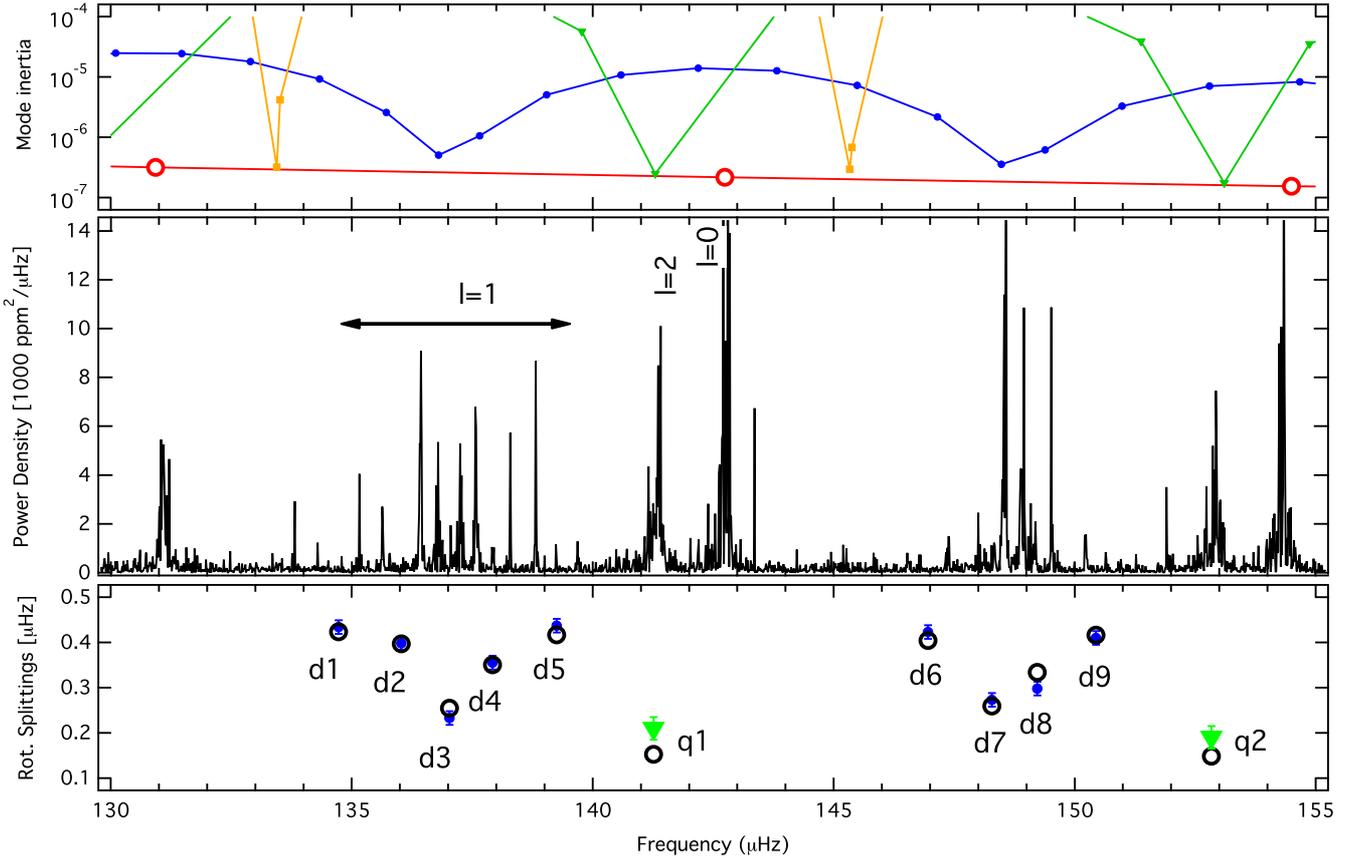}
\caption{Frequencies for the two central radial orders of KIC\,5006817. \new{Top panel:} The coloured dots connected with line segments indicate the inertia (right axis with arbitrary scale) of the $m=0$ adiabatic modes computed for a representative model for KIC\,5006817 (red circles, blue filled dots, green triangles, and yellow squares correspond to $\ell = 0, 1, 2$ and $3$ modes, respectively. 
\new{The middle panel provides a zoom on the frequencies in the range of the two central radial orders.}
Bottom panel: Measured rotational splittings for $\ell = 1$ (blue dots) and $2$ (green triangles) modes. Open black circles indicate the splittings that result from our best-fit 2-zone model. Dipole modes are labeled with $d$, quadrupole modes with $q$. \label{fig:modelPanoramix}}
\end{figure*}

We focused on the two radial orders (i.e. the frequency range of $\sim$130 to $\sim$155\,$\mu$Hz), in which 9 rotationally split dipole modes with good signal-to-noise were extracted. Additionally, we \new{tested if} rotational splittings from two $\ell = 2$ modes \new{can be measured}. In Fig.\,\ref{fig:modelPanoramix} we show the power density spectrum of the two radial orders of KIC\,5006817 
along with the mode inertia of the $\ell = 0$ to $3$ modes computed for our representative model, in the adiabatic approximation. 

\begin{figure}[t!]
\includegraphics[width=\hsize]{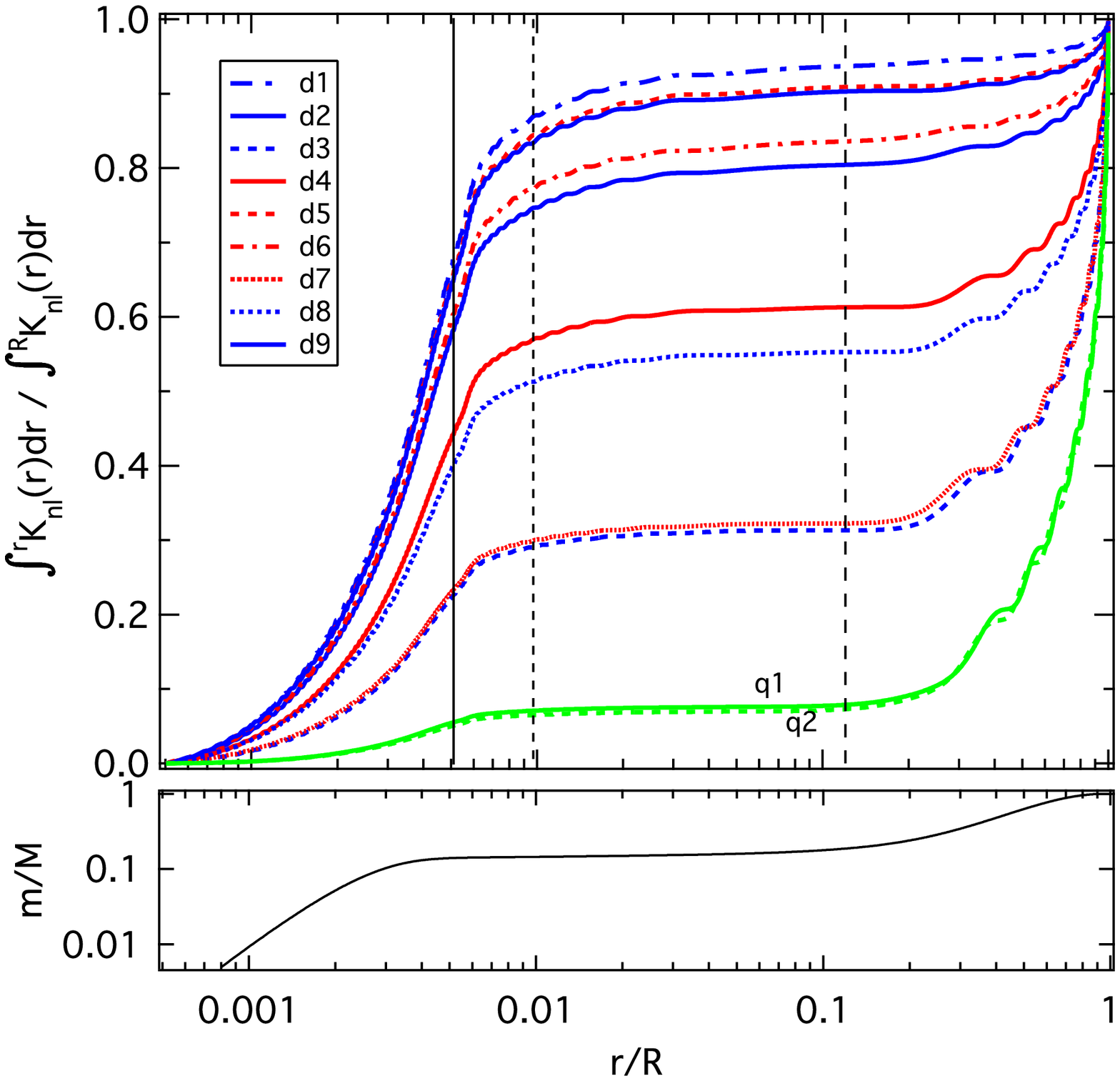}
\caption{Normalized \new{integrated} rotation kernels of the dipole $\ell=1$ (blue and red) and quadrupole $\ell=2$ modes (green) for the two central radial orders as a function of fractional radius. 
The kernel labels refer to modes in Fig.\,\ref{fig:modelPanoramix}. 
The vertical lines indicate from left to right the border of the He core (solid line), the H-burning shell (short dashed line), and the base of the convective envelope (long dashed line). Bottom panel: The model\new{'s} fractional mass as a function of the fractional radius. The labels (not the colours) of dipole and quadrupole modes are consistent with Fig.\,\ref{fig:modelPanoramix}.
\label{fig:rotationalKernels}}
\end{figure}

In order to determine which layers dominate the rotational splitting of a given mode, we computed the adiabatic eigenfunctions for the 9 dipole and 2 quadrupole 
modes \citep[using the nonradial nonadiabatic stellar pulsation code by ][]{guenther1994} as well as their rotational kernels.
 The integrated and normalized kernels are shown as a function of the fractional radius in 
Fig.\,\ref{fig:rotationalKernels}, illustrating that all dipole modes contain a significant contribution from the fast rotating core region, even the most p-mode dominated ones (indicated as d3 and d7). The kernels are almost flat in the radiative region of the envelope (between the two dashed vertical lines). This is in agreement with the analysis of the kernels for \object{KIC\,8366239} \citep{beck2012}. The dipole splittings provide an upper limit for the surface rotation rate. More promising are the $\ell = 2$ kernels, but even those contain significant contributions from the core and from the inner part of the convective envelope. 

For a better quantitative result, we deduce the rotation profile $\Omega(r)$ such that
\begin{equation} \label{EqRotPro}
\delta f_{n,\ell} = \frac{1}{2\pi I_{n,\ell}} \int_{r=0}^{R} K_{n,\ell}(r) \Omega(r) dr,
\end{equation}
is satisfied for the measured splittings. Here $I_{n,\ell}$ and $K_{n,\ell}$ are the mode inertia and the rotational kernel of a given mode, respectively, and $R$ is the radius of the model. To solve this equation, several inversion techniques have been developed in the past with the aim to determine the internal rotation profile of the Sun. A summary how to apply them to an evolved star is available in \citet{deheuvels2012}.

The inversion of the integral in Eq.\,(\ref{EqRotPro}) is a highly ill-conditioned problem and either requires numerical regularisation or localized averages of the true rotation profile in different regions of the star. We used both, adopting the regularized least squares method \citep[RLS; e.g.,][]{JCD1990} and the subtractive optimally localized averages technique \citep[SOLA; e.g.,][]{Schou1998}. In doing so, we can determine the core to surface rotation rate. However, as soon as we increased the number of shells, aiming to locate where the transition between fast rotating core and the slowly rotating envelope is taking place, both methods failed. Either the results became numerically unstable or it was  impossible to evaluate the reliability of the result.

We therefore chose to apply a Bayesian forward modelling approach. We computed synthetic rotational splittings for a model (using Eq.\,\ref{EqRotPro}), which was first divided into several solidly rotating shells, where the rotation rate of each shell was treated as a free parameter. To fit the synthetic splittings to 
the measured ones, we used a Bayesian nested sampling algorithm called {\sc MultiNest} \citep{feroz2009} that provides a probability density distribution (PDD) for each fitted parameter, from which we assessed the best-fit values and their uncertainties. The PDDs allowed us to test the reliability of a specific model, e.g.\ a flat PDD for a specific shell implies that the measured 
rotational splittings do not contain information about the rotation rate in this specific region of the star. Additionally, {\sc MultiNest} provides the model evidence, which allows \new{us} to compare different assumptions for the rotation rates in the shells and evaluate which one reproduces the observations best. We 
tested different numbers and combinations of shells and found that a 2-zone model (core and envelope) is better by several orders of magnitude. Even introducing regularisation priors (e.g., a smooth gradient between consecutive shells) did not improve the description of $\Omega(r)$.

\subsection{Core and surface rotation \label{sec:RotationRates}}
The final model we adopted is a 2-zone one with a core (inner 1 percent in radius or 14 percent in mass) and an envelope zone, where we found that the placement of the exact position of the border between the two zones only marginally influences the result. Our best fit constrained the rotation frequency (or period) for the core and the envelope to be $0.93\pm0.02\,\mu$Hz ($\sim$12.5\,d) and $0.07\,\pm0.02\mu$Hz ($\sim$165\,d), respectively. This corresponds to a core-to-surface rotation rate of about 13. The envelope rotation rate is an average value for 99 percent of the radius of the star and very likely overestimates the true surface rotation. This asteroseismic result points towards a surface rotation velocity of 1.9\,km\,s$^{-1}$. 

Including the two $\ell=2$ modes only marginally affects the rotation rate but increases the uncertainty, which is why they were not taken into account for the final result. We conclude that the currently available observations do not allow us to assess the detailed rotational gradient in the envelope of KIC\,5006817, as it was possible for the Sun. We are limited to describing the rotational profile of red giants with a step function.

\subsection{Inclination of the axis of rotation and pulsation}
The inclination angle of the rotation axis towards the observer can be deduced from the rotational splittings. \new{For solar-like oscillators, the excitation of all $2\cdot\ell+1$ components and the equipartition of energy is assumed if the time base of the observations resolves the lifetime effects of a mode.}

\new{For dipole modes, the inclination is determined by the height of the $m$=0 and $m$=$\pm$1 modes,} provided that all components \new{with the same $|m|$} are excited such that they have the same height in power density \citep[][]{gizon2003, ballot2006}. We have also shown that each component of a split multiplet has a slightly different cavity which should result in slightly different heights and lifetimes, an effect absent in the Sun as it has no mixed modes. To compensate for these differences, we do not fit the heights of the $m=\pm1$ in a given mode individually, but force the heights of the fit to be equal. 

Simultaneously fitting all dipole modes in at least one radial order  compensates in principle the effects of a changing \new{level} of \new{mixed character} between p and g modes. This transition is symmetrically mirrored around the pure dipole pressure mode. We tested this approach for the radial orders around 135 and 150\,$\mu$Hz, resulting in inclination values of i$_{\rm rot}$=73$^\circ\pm$3$^\circ$ and i$_{\rm rot}$=80$^\circ\pm$3$^\circ$ respectively. 
\new{A global fit of both radial orders lead to i$_{\rm rot}^{\rm global}$=76$^\circ\pm$4$^\circ$, assuming alignment of the rotation and pulsation axis.}
\new{We compare these values to the mean inclination of i$_{\rm rot}^{\rm mean}$=77$\pm$9, obtained from the individual rotational split multiplets of the pressure-dominated modes in this frequency range (Tab.\,\ref{tab:individualInclinations}). These modes have the highest signal and the shortest lifetimes, and therefore are closest to the assumption of equipartition of mode energy. }
\newer{The} uncertainty is \newer{an} underestimate as it was computed adopting the assumptions mentioned above.

\begin{table}[t!]
\caption{\label{tab:individualInclinations}
\new{Inclination values for individual pressure dominated modes.}}
\centering
\begin{tabular}{cc}
\hline\hline
$m$=0	& 	inclination \\\hline
137.02	&	70.2$\pm$5 \\
137.93	&	76.3$\pm$5 \\
148.29	&	82.9$\pm$4 \\
149.19	&	80.2$\pm$4 \\\hline
mean & 77$\pm$9\\\hline
\end{tabular}
\tablefoot{The individual inclinations for the pressure dominated dipole modes, centred on the $m$=0 component, in the range between 130 and 155\,$\mu$Hz (Fig.\,\ref{fig:modelPanoramix}).}
\end{table}%

 \section{Spectroscopy of KIC\,5006817
 \label{sec:Spectroscopy}}
 \begin{table}[t!]
 \tabcolsep=4pt

 \begin{center}
 \caption{Fundamental parameters of the red giant KIC\,5006817.
  \label{tab:specpar} }
 \begin{tabular}{cccccc}
 \hline\hline
 T$_{\rm eff}$ $[K]$& $\log{g}$[dex] &  [Fe/H]\,[dex] &  $v_{\rm
 micro}$\,[km/s]  & LB\,[km/s]  \\
 \hline
 5000 $\pm$250 & 3.0 $\pm$0.5 & -0.06 $\pm$0.12 & 3.0 $\pm$0.5 & 8 $\pm$1  \\
 \hline
 \end{tabular}
 \tablefoot{Results for the effective temperature $T_{\rm eff}$, surface gravity $\log
 g$, and microturbulence $\nu_{\rm micro}$ of the red giant component. \new{The last column gives the total line broadening as defined in Section \ref{sec:fundParamSpec}}.}
 \end{center}
 \end{table}

 To obtain an independent estimate of the eccentricity of the system,
 KIC\,5006817 has been monitored spectroscopically with the \textsc{Hermes} spectrograph
(cf. Section \ref{sec:Observations}) in 2012 \new{and 2013}. The 60 observations span about 160\,d,
 during which the periastron passages were monitored with several
 observations a night. The radial velocities derived from these
 spectra are shown in the top panel of Fig. \ref{fig:RV}.

 \subsection{Fundamental parameters \label{sec:fundParamSpec}}
 The first 44 individual spectra were shifted by the derived radial velocity value and averaged to produce a
 high signal-to-noise ratio (S/N) spectrum to determine the stellar atmospheric
 parameters like effective temperature, T$_{\rm eff}$, surface gravity
 $\log g$, microturbulence $\nu_{\rm micro}$ and the \new{total line broadening from  rotation and macroturbulence} ($v \sin i + v_{\rm macro}$). 
 \new{Given the seismic estimate of rotation and inclination, the total line broadening is dominated by macroturbulence.}
 We used the combination of local thermal equilibrium
 (LTE) Kurucz-Castelli atmosphere models \citep{castelli04} with the
 LTE abundance calculation routine MOOG by \citet{sneden73}. A
 detailed description of all steps needed to derive the atmospheric
 parameters can be found in, e.g, \citet{desmedt12}.

 The determination  of the atmospheric parameters was based upon Fe I
 and Fe II lines which are abundantly present in red giants spectra. The
 Fe lines used were taken from VALD linelists \citep{VALD}. We
 first calculated in MOOG the theoretical equivalent width (EW) of all
 available Fe I and Fe II in a wavelength range between 400 and
 700\,nm~for a grid, centred on the seismic parameters found
 (Table\,\ref{tab:asteroseismicValues}). The equivalent widths (EW) of
 lines were measured via direct integration, the  abundance of the
 line was then computed by an iterative process where theoretically
 calculated EWs were matched to the observed EW. If the
 calculated EW deviated from the theoretical EW by a factor of 10, the
 line was rejected due to possible blends. The stellar parameters are
 listed in Table \ref{tab:specpar} and are based on the results from 52 Fe I and 32 Fe
 II lines.

 \subsection{Spectral disentangling}
 In order to look for signatures of a companion in our spectra, we
 used the spectral disentangling ({\sc spd}) method as implemented in
 the {\sc fdbinary} code \citep[][]{Ilijic2004}. Being applied in
 Fourier space, the method provides a self-consistent and fast
 solution for the individual spectra of stellar components of a
 multiple system and a set of orbital parameters simultaneously
 \citep[][]{Hadrava1995}. The {\sc spd} method usually requires a
 time-series of spectra with good, homogeneous orbital phase coverage
 and delivers high S/N mean spectra of individual components.

 \begin{figure*}[ht!]
\hfill{}
\includegraphics[width=0.8\hsize]{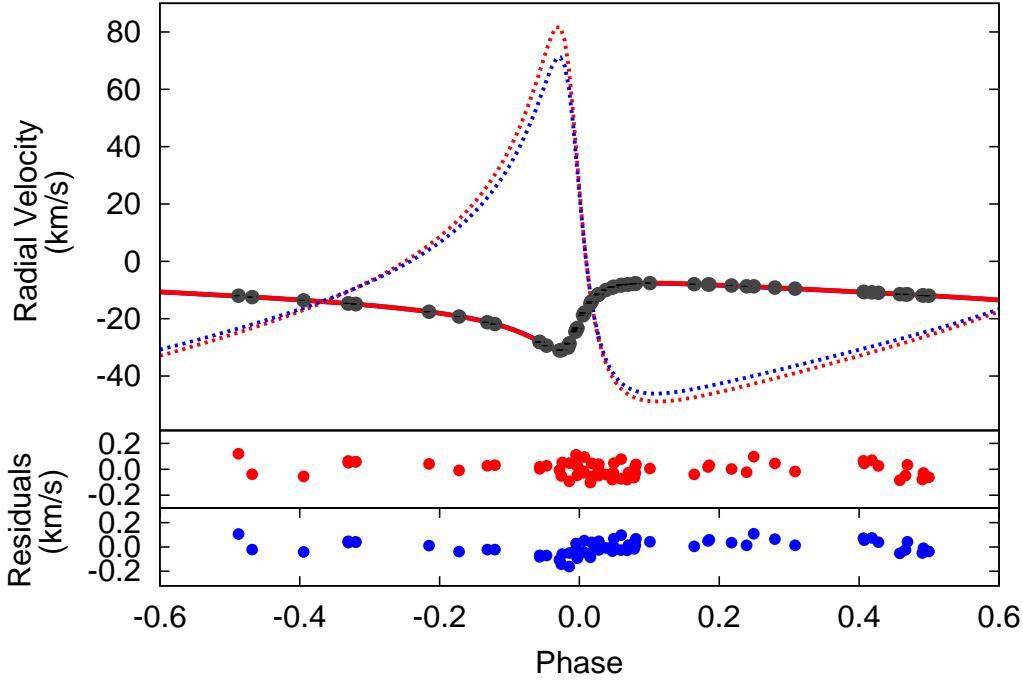}
\hfill{}
\caption{Radial velocity observations (gray points) with the best-fit model without beaming (red solid line) and with beaming (blue solid line - not visible) phased over 94.82\,d. The uncertainty in the radial velocity is smaller than the size of the points. The model with beaming cannot be seen on this scale due to the overlap of the model without beaming. The model of the secondary component is presented for both the beaming (blue dashed line) and non-beaming (red dashed line) best-fit models. Middle panel: the best-fit residuals of the binary model without beaming (red points). Lower panel: the residuals of the best-fit binary model with beaming (blue points).}
\label{fig:RV}
\end{figure*}

 We used the highest S/N spectra to perform the spectral
 disentangling in a wide wavelength range from 480 to 580\,nm. The
 spectra were split into smaller chunks, typically 5\,nm wide, to
 avoid strong undulations in the continuum of the resulting decomposed
 spectra from which the Fourier-based {\sc spd} method is known to
 suffer \citep[see, e.g.,][]{Hadrava1995,Ilijic2004}. Besides the above
 mentioned metal lines region that also contains the H$_{\beta}$ spectral line, we
 also focused on a few other regions centered at some helium lines
 (He~I 447, 492, and 502\,nm, and He~II 469,
 541, and 656\,nm,) as well as the H$_{\alpha}$ line. This was done to verify
 whether the companion could be a white dwarf showing helium and/or hydrogen
 lines only. In neither of these regions could we detect the lines of the
 secondary. We estimate the detection limit of the order of 3 percent of
 the continuum, which means that any contribution below this level
 would not be detectable in our rather low S/N spectra ($\simeq$20 - 30 in Johnson V).

 We also attempted to go beyond our actual detection limit of ~3
 percent, by applying the least-squares deconvolution (LSD) method
 \citep[][]{Donati1997} to the 44 individual spectra. This method is based on
 the two fundamental assumptions of self-similarity of all spectral lines and
 linear addition of blends, and allows to compute a high-quality average line
 profile, which is formally characterized by a very high S/N. The first
 assumption requires hydrogen and helium lines as well as the metal lines with
 pronounced damping wings to be excluded from the calculations. Moreover, for
 slowly rotating stars (the case of KIC\,5006817), where the rotation is not the
 dominant source of the line broadening, the selfsimilarity is only applicable
 to the lines of (nearly) the same strengths. To account for this, we introduced
 a multiprofile technique as described by \citet[][]{Kochukhov2010}, which
 allows the computation of several average profiles simultaneously for several sets of
 spectral lines grouped, e.g., according to their relative strengths. The model
 is then represented as a convolution of the computed mean profiles with the
 {\it line mask} which contains information about the position of individual
 lines as well as their relative strengths.

 Furthermore, part of the lines in
 the spectrum (e.g., those with overlapping absorption coefficients) add up non-linearly which requires a revision of the second fundamental assumption of the
 technique. In order to account for the model imperfections due to non-linear
 blending of the lines, we additionally fitted strengths of the individual lines
 from the mask to match the observations. This improved procedure \citep[for
 more details, see][]{Tkachenko2013} provided us with high S/N (of the order of
 450-500) LSD model spectra which we then used for the {\sc spd} in three metal
 lines regions. Similar to our previous experience with the original spectra, we got a null result, in the sense
 that no signature of the secondary has been detected, this time, in the high
 S/N composite spectra. In this case, our detection limit is estimated to be of
 the order of 1 percent of the continuum, which is mainly due to the imperfect
 continuum normalization rather than the observational noise. We thus conclude
 that KIC\,5006817 is a single-lined spectroscopic binary.

\begin{figure}[t!]
\includegraphics[width=\hsize]{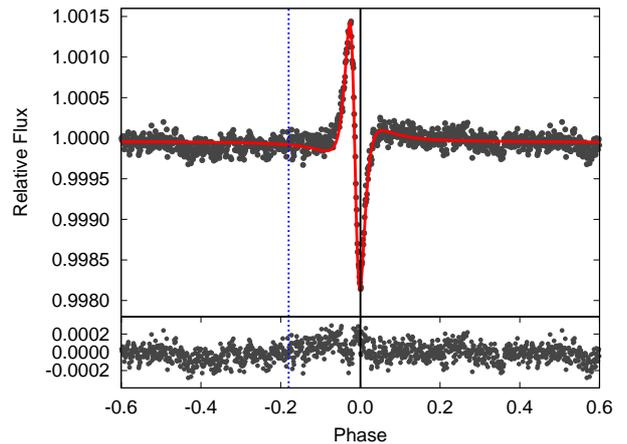}
\caption{Theoretical {\sc phoebe} model without beaming (red line) and observed 
light curve (black points) of the phased (94.82\,d), binned \textit{Kepler} long cadence data of Quarters 0--12. Lower 
panel: the residuals of the best-fit model. The dashed and solid lines are centred on the times of superior and inferior conjunction 
respectively.}
\label{fig:LC}
\end{figure}
%
%

\section{Binary parameters of KIC\,5006817\label{sec:OrbitalStudy}}

\new{The morphology of the photometric light curve is a consequence of ellipsoidal modulation in an eccentric system. The shape of the ``heartbeat" feature in the light curve is a function of the inclination, which dictates the peak to dip ratio; the eccentricity, which affects the relative width of the feature; and the argument of periastron which affects the symmetry of the feature. 
The magnitude of the heartbeat feature is dependent on the radii of the components, their masses and gravity darkening exponents, and the orbital inclination.} 

\new{To study the binary properties of KIC\,5006817, we simultaneously modelled the photometric and radial velocity data of KIC\,5006817 using the modelling code, {\sc phoebe} \citep{Prsa2005}.} This code is an extension of the Wilson-Devinney code \citep{Wilson1971,Wilson1979,Wilson2004} and combines the complete treatment of the Roche potential with a detailed treatment of surface and horizon effects such as limb darkening, gravity darkening, ellipsoidal modulation and reflection effects to arrive at a comprehensive set of stellar and orbital parameters.

The orbital period of the binary is close to the length of one \textit{Kepler} quarter ($\sim 3$ months), which is the typical time scale of long term instrumental trends in \textit{Kepler} data. Consequently, in the case of KIC\,5006817, accurate detrending of the light curve is a challenge. We have therefore manually extracted each light curve, on a Quarter-by-Quarter basis, from pixel level data to create the best behaved light curve possible. We then fitted and divided out a linear trend from each Quarter to detrend and normalise the data. We selected a linear trend to avoid removing the beaming information from the light curve.

\subsection{Orbital ephemeris}

From the long cadence \textit{Kepler} photometry of KIC\,5006817 (Quarters 0--12) we determined the zero point in the data (the time of the periastron minimum) to be 2455019.221\,$\pm$\,0.008 using the {\sc kephem} software package \citep{Prsa2011}. Combining this with the spectroscopically determined period (94.812\,$\pm$\,0.002\,d) we obtained the following ephemeris in the Barycentric Julian date:

{\rm Min I} = 2455019.221\,$\pm$\,0.008 + 94.812\,$\pm$\,0.002\,d $\times$ E, 

\noindent
where $E$ is the number of orbits. 
\subsection{Input parameters}

In our fits, we have fixed the effective temperature of the primary component to the spectroscopic value: $T_{\rm eff,1}=5000\,$K (Table\,\ref{tab:specpar}). 

Assuming the secondary component is a main sequence star (from its lack of visibility in the spectra), we can place an upper limit of $\sim 5400$\,K on its effective temperature by assuming that it contributes 1 percent of the total flux (the lower limit for spectral disentangling). This temperature is an upper limit and is likely an overestimate, given the mass determined from the binary model ($\sim$0.3\,$M\sun$, see Section\,\ref{sec:massRatio} and Table\,\ref{tab:ParamBestFitModel}). Due to the evolutionary state of the primary and low mass of the secondary, we assumed that both components have substantial convective outer envelopes and consequently adopted the standard albedo value of 0.6 \citep{Rucinski1969b,Rucinski1969a}. As both objects radiate towards the infrared end of the optical spectrum, as suggested by \citet{Diaz-Cordoves1992} and \citet{vanHamme1993}, we have selected the square-root limb darkening law \citep[Eq.\,6 in][]{Diaz-Cordoves1992}. 

\subsection{Fitting procedure}
%
\begin{table}[t!]
\hfill{}
\caption{
\label{tab:ParamBestFitModel}
Parameters and coefficients for the {\sc phoebe} best-fit model to the \textit{Kepler} light curve for  Quarters\,0$-$12 long cadence data for the non-beaming and beaming cases. For the beaming case we assumed 100\% of the flux comes from the primary component.}
\begin{tabular}{l|r|r}
\hline\hline
Parameter   &{Non-beaming} & Beaming\\
\hline
Mass ratio, $q$					& 0.199$\pm$0.001 	& 0.20$\pm$0.03  \\
Secondary mass ($M\sun$), $M_2$	& 0.30$\pm$0.01   	& 0.29$\pm$0.03\\
Semi-major axis ($R$\sun), $a$  		& 106.1$\pm$0.5   	& 105.6$\pm$ 0.9\\
Orbital eccentricity, $e$			& 0.71$\pm$0.01   	& 0.71$\pm$0.02\\
Argument of periastron (rad), $\omega$ & 4.0$\pm$0.1   	& 4.01$\pm$0.06  \\
Orbital inclination ($\degrees$), $i$	 	& 62$\pm$4    		& 61$\pm$6  \\
Primary potential, $\Omega$$_1$  	& 16.7$\pm$0.2  	& 16.2$\pm$0.2  \\ 
Gamma velocity (km\,s$^{-1}$), $\gamma$ 	& -14.01$\pm$0.01 & -14.43$\pm$0.08  \\
Primary $\log$\,$g$ (cgs),  $\log$\,$g$$_{1}$ & 3.078 $\pm$0.007 & 3.022$\pm$0.007  \\
\new {Gravity darkening exponent, {\sc grd}}		& \new {1.00$\pm$0.03} & \new {1.07$\pm$0.03}\\
Primary fractional point radius                         	& 0.0644          		& 0.0679  \\
Primary fractional pole radius                          	& 0.0637          		& 0.0676  \\
Phase of periastron				& 0.0121          		& 0.0094  \\
Primary $x_1$ coeff.		                	& \new{0.718}           		& 0.717  \\
Primary $y_1$ coeff. 	                        	& \new{0.716}           		& 0.714  \\
\hline
\multicolumn{1}{l}{Fixed Parameters}&\multicolumn{2}{|c}{Values: Both Cases}\\
\hline
Primary $T_{eff}$ (K)            &\multicolumn{2}{|c} {5000$\pm$250}\\
Third light                     &\multicolumn{2}{|c} {0.0}\\
Orbital Period (d)           	&\multicolumn{2}{|c} \new{ {94.812$\pm$0.002}}\\
Time of primary minimum (BJD) 	&\multicolumn{2}{|c} {245019.221$\pm$0.008}\\
Primary Bolometric albedo       &\multicolumn{2}{|c} {0.6}\\
\hline
\end{tabular}
\tablefoot{The secondary component's potential, radius and $\log$\,$g$ are not noted as these parameters have no signature in the light curve and radial velocity data. The fractional radii quoted are the radii relative to the semi-major axis. For the mass and radius of the primary component, see Table\,\ref{tab:asteroseismicValues}. The limb darkening coefficients ($x_1$ and $y_1$) are for the square root limb darkening law and were taken from the {\sc phoebe} limb darkening tables \citep{Prsa2011}. }
\\
\end{table}

After manually tweaking the initial parameters until an approximate fit to 
the photometric data and the spectroscopic radial velocities was accomplished, we applied differential corrections \citep{Wilson1976} to optimize the parameters.

The binary model created has the mass and radius of the primary component fixed to the asteroseismic value. Using differential corrections we simultaneously fitted the eccentricity and argument of periastron to the radil velocity and light curves; fitted the semi-major axis and gamma velocity to the radial velocity curve and fitted the remaining parameters to the light curve only. The fitted and fixed parameters of the best-fit model are listed in Table\,\ref{tab:ParamBestFitModel}, Figs.~\ref{fig:RV} and \ref{fig:LC} show the phase binned light curve and radial velocity data with the best-fit model (red line) in the upper panel and the corresponding residuals in the lower panel.

\subsection{The puzzling absence of a Doppler beaming signal}
\begin{figure}[t!]
\hfill{}
\centering
\includegraphics[width=\hsize]{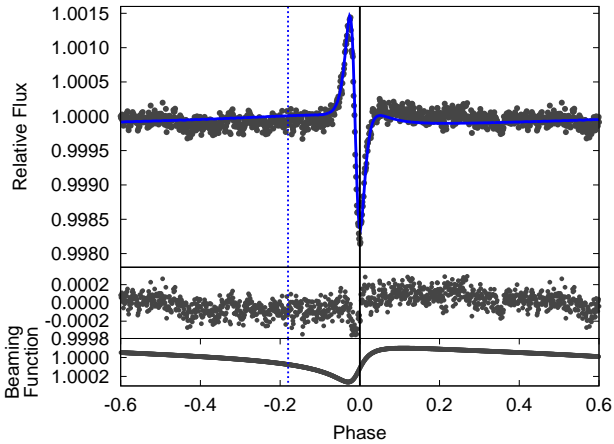}
\caption{\label{fig:LCbeam} Theoretical {\sc phoebe} model similar to Fig.\,\ref{fig:LC}, but including beaming (blue line). The dotted and dashed lines are centred on the times of superior and inferior conjunction, respectively.
Middle panel: the residuals of the best-fit beaming model. Lower panel: the Doppler beaming function that has been added to the model to incorporate beaming. The axis is inverted for comparison with the residuals.  
}
\hfill{}
\end{figure}

So far, we have not taken into account the effect of Doppler beaming on the light curve. Doppler beaming is caused by the radial velocity of the two stars and is the combined effect of shifting the stars' spectral energy distributions with respect to the \textit{Kepler}\ bandpass, aberration and an altered photon arrival rate. The net result of Doppler beaming is an increase in the observed flux from a star when it moves towards the observer, and a decrease when it moves away from the observer. It was predicted to be seen in \textit{Kepler}\ data by \citet{LoebGaudi2003} and \citet{ZuckerMazeh2007}, and has recently been observed in several systems from ground-based data as well as \textit{Kepler}\ and CoRoT light curves \citep[see e.g.][]{MazehFaigler2010,van-KerkwijkRappaport2010,ShporerKaplan2010,BloemenMarsh2011}. 

Based on the radial velocity measurements of the primary star (semi-amplitude, $K$\,=\,11.78 \,kms$^{-1}$ - see Table\,\ref{tab:sampleOrbits}), we estimate that Doppler beaming should have a significant contribution in the light from the red giant in KIC\,5006817, of the order of $\sim$300 ppm. Since spectroscopy indicates that the secondary component in the binary has an insignificant luminosity compared to the primary, we looked into the effect of Doppler beaming assuming that all the observed flux is emitted by the red giant primary component. The Doppler beaming signal was modelled following Eq.\,(2) in \citet{BloemenMarsh2011}. The Doppler beaming coefficient of the red giant primary, which takes into account the spectrum of the star and the wavelength of the observations, was computed using Eq.\,(3) of \citet{BloemenMarsh2011} to be $\left<B_1\right>= 4.59 \pm 0.21$ from Kurucz 2004 model spectra \citep{Castelli2004}.

Figure~\ref{fig:LCbeam} depicts the phase-folded light curve and best-fit model including beaming for KIC\,5006817. The residuals of this model, however, contain a significant sinusoidal wave, similar in nature to that of the beaming function. The beaming signal thus seems to be invisible in the observed light curve. At phase zero the beaming signal has been fitted by adjusting the inclination of the model, however, the fit is still less adequate in this region than for the non-beaming case. 

Possible reasons for the absence of a beaming signal in the data are a very high third light contamination which reduces the observed amplitude of the effect; a significant beaming signal from the secondary that partly or fully cancels out the beaming signal of the primary; or that the inherent long period instrumental trends in the \textit{Kepler} data are concealing the beaming signal. The first possibility can be ruled out, since no bright source of third light is seen close to the target and no contaminating light was found in the spectra. To evaluate the second possibility - that the beaming signal from the secondary is cancelling out the primary beaming signal, we scanned the parameter space to find the light ratio (with beaming from both components) that best fitted the data. We only fitted the continuum section of the light curve, since the beaming effect can be compensated for at the phase of the periastron variation by adjusting the inclination.
\begin{figure}[t!]
\hfill{}
\centering
\includegraphics[width=\hsize]{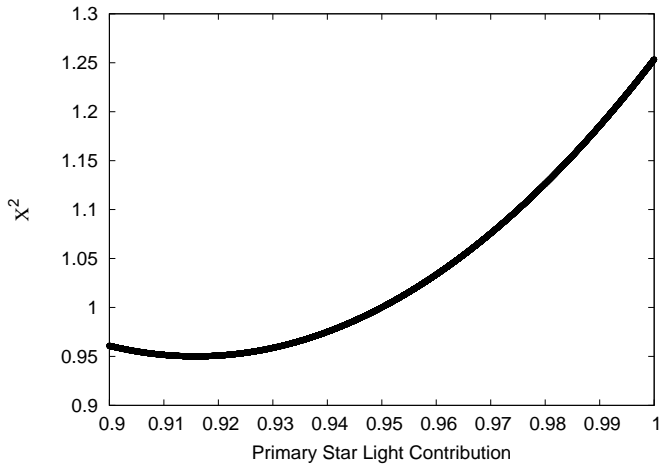}
\caption{The \chisq goodness of fit to the continuum light curve (light curve with the periastron variation removed) as a function of the fraction of light contributed by the primary component, for multiple models. For each model the fraction of light was randomised (90--100 percent for the primary component) and the beaming function was calculated for both the primary and secondary component. The goodness of fit was then assessed away from the periastron variation as only the beaming has an effect on this section of the light curve.}
\label{fig:light_cont}
\hfill{}
\end{figure}
When including Doppler beaming for the secondary component, we assumed that it is a main sequence object with a mass of 0.3\,$\Msun$, giving a Doppler beaming coefficient of $\left<B_2\right> \simeq 6.5$. 

\begin{figure}[t!]
\includegraphics[width=\hsize]{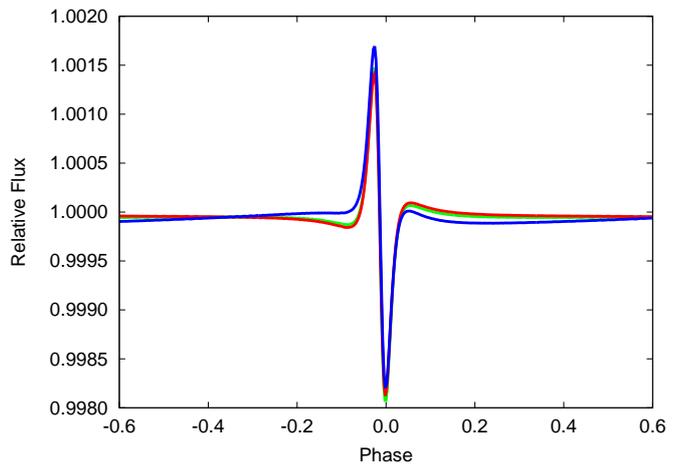}
\caption{The best-fit model to the phased light curve data (red), the same model with beaming and 100 percent light contribution from the primary component (blue), and with beaming and 91.5 percent light contribution from the primary component (green) \new{- the best fit model when allowing the flux ratio to be fitted}.}
\label{fig:beam_comp}
\end{figure}

Figure~\ref{fig:light_cont} shows the $\chi^2$ value as a function of primary light contribution. We find a clear preference for a contribution of 91.5 percent from the primary component and 8.5 percent from the secondary component. As shown in Figure\,\ref{fig:beam_comp}, the modelled light curve for the preferred light ratio (green line) is approximately equal to the light curve excluding beaming, suggesting that the preferred configuration leads to the secondary essentially cancelling out the beaming from the primary component. If the secondary is a 0.3\,$\Msun$ main sequence star, it can not, however, contribute on the order of 8 percent of the light in the system, and furthermore, a main sequence star this bright would have been easily detected in our spectroscopic data. We also considered the option that the secondary is a continuum white dwarf as this would not necessarily show up in the spectra. However, we can rule out this possibility as the temperature would need to be greater than 40,000\,K for it to contribute 8.5 percent of the flux and as such the light curve would show an extreme reflection effect (the white dwarf reflecting off the red giant), which we do not observe. We would also see evidence for such a white dwarf in the spectroscopic data, as continuum white dwarfs (which do not have a signal in spectroscopic data) have effective temperatures $\lesssim$\,12\,000\,K. Furthermore, a hot star would have a lower beaming factor and thus require a greater flux contribution from the secondary component.

\new{Other than missing or incomplete physics in the binary models,} the final possibility that we have postulated is that the \textit{Kepler} satellite is not stable enough on longer timescales to preserve the beaming signal in the data. While this seems like a more convincing option, given that the trends in the \emph{Kepler} light curves have a larger amplitude than the beaming signal, and that the timescales of the orbital period and a \emph{Kepler} quarter are similar, we find it surprising that we are unable to detect any signature of the beaming signal given that we used a minimally evasive detrending method. We also note that the beaming signal was still missing when using a higher order Legendre polynomial in place of the linear trend. 

The slope of the beaming signal is largest at the phase of the periastron brightening, and therefore significantly influences the optimal set of parameters found when fitting the heartbeat. Since we do not understand the absence of the beaming signal, we present the results of the fits to the data both including beaming, assuming 100 percent light contribution from the primary, and without beaming. The optimal parameter values for both cases are listed in Table\,\ref{tab:ParamBestFitModel}.

\subsection{Mass ratio, primary potential and gravity darkening degeneracies \label{sec:massRatio}}

\begin{figure}[t!]
\includegraphics[width=\hsize]{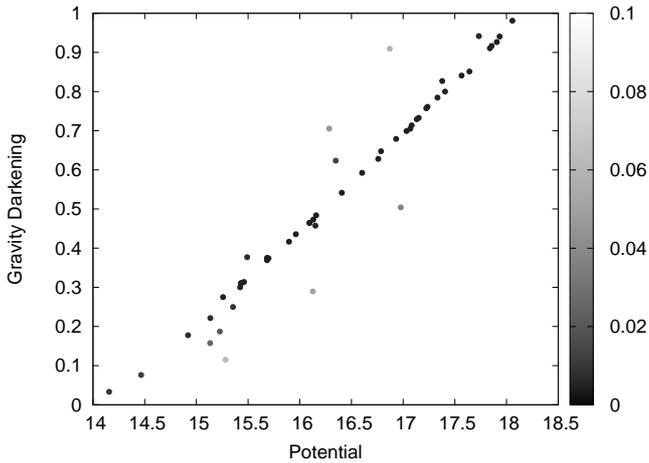}
\caption{The gravity darkening value as a function of the primary potential, where each point represents an individual binary model. For each model the gravity darkening exponent was determined randomly (between 0.0 and 1.0) and a model generated by fitting the light factor and potential, whilst keeping all other parameters fixed. The points are coloured by the \chisq goodness of fit to demonstrate that the outliers are a consequence of an inadequate fit (the lower \chisq values denote the better models, here depicted in black).}
\label{fig:grbVpot}
\end{figure}

\new{When fitting the binary characteristics, we found that the mass ratio, primary potential (which is essentially the inverse of the primary radius) and gravity darkening exponent are degenerate with each other. To assess the level of degeneracy between the potential and the gravity darkening exponent}, a scan of the parameter space was undertaken whereby the primary gravity darkening exponent was randomly adjusted following which the primary potential and light factor were fitted to the light curve using differential corrections. Figure~\ref{fig:grbVpot} shows the gravity darkening value and corresponding potentials for multiple models. The points are coloured with respect to their \chisq\ value to show the goodness of fit for each individual model. The results show a complete degeneracy between the primary potential and primary gravity darkening exponent. \new{Repeating this experiment for the primary potential and mass ratio, we again found a complete degeneracy. For this reason we elected to fix the primary mass and radius to the asteroseismically determined values, 1.49\,$\pm$\,0.06\,$\Msun$ and 5.84\,$\pm$\,0.08\,$\Rsun$, respectively.}

\begin{figure}[t!]
\centering
\includegraphics[height=150mm]{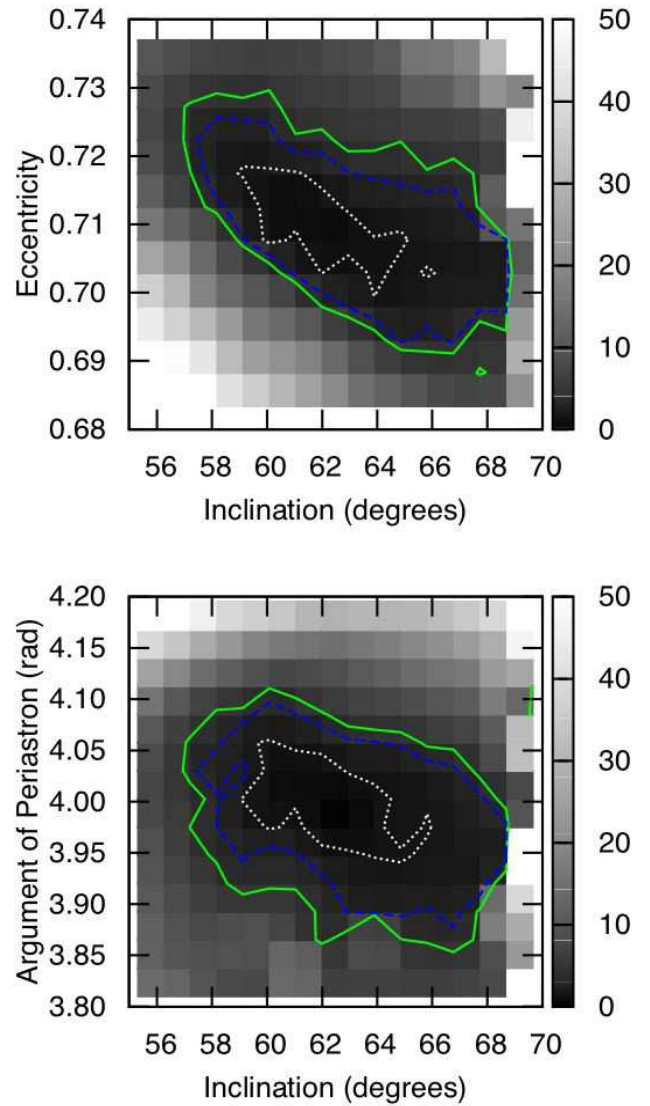}
\caption{\label{fig:MC} Density maps of the Monte Carlo simulations of the inclination vs.\ eccentricity (upper panel) and the inclination vs.\ argument of periastron (lower panel) for the non-beaming case. The grey scale depict the \chisq values mapped across the grid. The colour bar depicts the grey scale pertaining to the different values of $\chi^2$. The contours (from inner to outer) denote the 1$\sigma$ (white dotted line), 2$\sigma$ (blue dashed line) and 3$\sigma$ (green solid line) confidence intervals. The uncertainties for the inclination, argument of periastron and eccentricity were determined using the 1$\sigma$ confidence contours displayed.}
\end{figure}

Keeping the gravity darkening exponent as a free parameter, we found a best binary model fit to the gravity darkening coefficient to be 1.00\,$\pm$\,0.03 and $1.07 \pm 0.03$ for the beaming and non-beaming cases, respectively. \new{These values are not in agreement with the accepted value of {\sc grd}\,=\,0.32 for a star with a convective envelope \citep{Lucy1967}, although this is an empirically determined value for main sequence objects. More recent literature suggests an increased value of $\sim$0.5 based on computations of atmosphere models \citep{ClaretBloemen2011}. In this case the value is specific for stars with an effective temperature and surface gravity close to that of the red giant in KIC\,5006817. While this value is closer to that determined, there is still a large discrepancy between the observed and theoretical values. A possibility is that the uncertainties of the asteroseismic mass and radius are underestimated. However, a closer look suggests that the radius would have to increase by three sigma and the mass decrease by three sigma to reach the gravity darkening exponent suggested by \citet{ClaretBloemen2011}. While a three sigma limit may be plausible, the change in values would require the density - the most constrained asteroseismic value - to deviate significantly from that determined, which is unlikely. The most likely explanation is that the accepted gravity darkening exponent needs to be revised or completely mitigated from the models \citep{Espinosa2012}.}

\subsection{Uncertainty Determination}

The uncertainties of the parameters were determined using two methods: through 
standard errors and their propagation, and through Monte Carlo heuristic scanning. A scan of the parameter space was undertaken for the most correlated (but not completely degenerate) parameters, which were determined by applying the correlation matrix function in {\sc phoebe}. With a fixed mass ratio and gravity darkening, the most correlated parameters were determined to be the inclination, eccentricity and argument of periastron. The magnitude of the periastron brightening was not found to be significantly correlated with these parameters, which determine the shape of the periastron brightening.

We applied Monte Carlo simulations to perturb the solutions of the 
eccentricity, argument of periastron and inclination. The applied method required the computation of the potential and phase shift, and the iterative randomisation of the eccentricity and argument of periastron by 10 percent, and the inclination by 20 percent. At each iteration a comparison between the model and phased data was made using the $\chi^2$ statistical test. The $\chi^2$ values for each solution were then mapped out across a parameter grid with confidence intervals, which serve as uncertainty estimates (cf.\ Fig.\,\ref{fig:MC}). The optimum combination of the displayed parameters can be identified from the density maps, where the 1$\sigma$, 2$\sigma$ and 3$\sigma$ uncertainty values are presented as contours.

\section{Combined asteroseismic and binary interpretation \label{sec:ResultComparison}}

\begin{figure}[t!]
\includegraphics[width=\hsize]{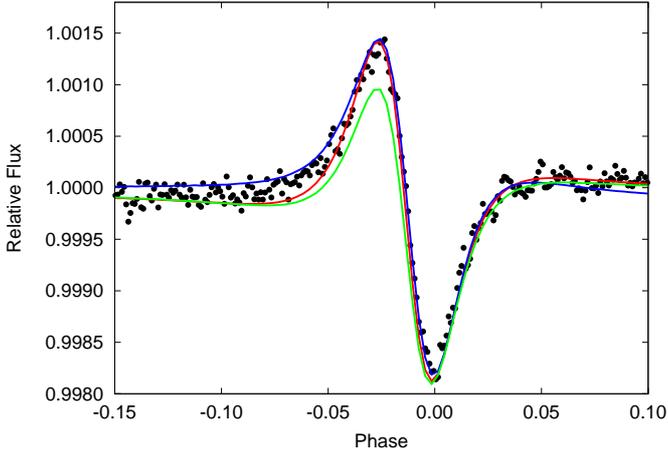}
\caption{Comparison between the data (grey points), the best-fit light curve model without beaming (red dotted line), the best-fit model with beaming (blue dashed line), and the best-fit model for an imposed orbital inclination of 76$\degs$, as determined through asteroseismology (green solid line). A change in the inclination changes the ratio between the maximum and minimum of the periastron variation (also known as the heartbeat event).}
\label{fig:incl-fit}
\end{figure}

Assuming the equilibrium tide model \citep{Zahn1966,Zahn1989,Remus2012}, which applies to stars with convective outer envelopes, we calculated approximate timescales for the synchronization and circularization of KIC\,5006817, using the work of \citet{Zahn1977}. The synchronization timescale denotes the amount of time needed for the star's rotational angular velocity to equal its orbital angular velocity at periastron (pseudo-synchronous rotation) and for the stars' axes to become perpendicular to the orbit. The circularization time scale is the maximum time the orbit will take to circularize. \new{Applying Eqn. 6.1 of \citet{Zahn1977} we} determined the synchronization timescale \new{due to gravitational interactions} to be 2\,$\times$\,10$^{12}$\,yr, \new{which is prohibitively large.} 

To determine whether KIC\,5006817 is synchronized \new{we compared the inclination of the orbit and the rotational axis, and considered} the rotation rate of the red giant. The inclination measured from rotational splitting is sensitive to the orientation of the rotational axis of the primary pulsator, while the inclination determined from the shape of the heartbeat event describes the orientation of the orbital plane. In the case of KIC\,5006817, these two inclination angles were found \new{to} agree within 2$\sigma$: $i_{\rm rot} = 77^\circ\pm9^\circ$ versus $i_{\rm orbit} = 62^\circ\pm4^\circ$ (61$^\circ$$\pm$6$^\circ$ for the beaming case). 

To assess this difference graphically, Fig.\,\ref{fig:incl-fit} compares the best-fit binary model without beaming (red) and with beaming (blue) and with the orbital model with an imposed orbital inclinations of 76$\degs$ (green) taken from asteroseismology. It is clear that the maximum in the light curve is not well approximated with such a high orbital inclination. On the other hand, the very low visibility of the zonal modes ($|m|$=0) in the center of the rotationally split triplets, under the assumption of equal intrinsic amplitudes, definitely excludes inclinations below 70$^\circ$. 

If the system were rotating pseudo-synchronously, following equation (44) of \cite{hut1981}, using the observationally derived eccentricity and orbital period, one would expect a rotation period of about 11\,d for KIC\,5006817. Such rapid surface rotation is immediately ruled out from the width of the absorption lines in the spectrum of the primary. Moreover, the splitting of the $l$=1 modes points towards a rotation period of at least 165\,d, which is more than 1.7 times the orbital period. Thus we conclude that, as expected, KIC\,5006817 is not in a synchronized orbit.

\section{The impact of stellar evolution on eccentric binary systems \label{sec:sdBDiscussion}}
\begin{figure}
\centering
\includegraphics[height=110mm]{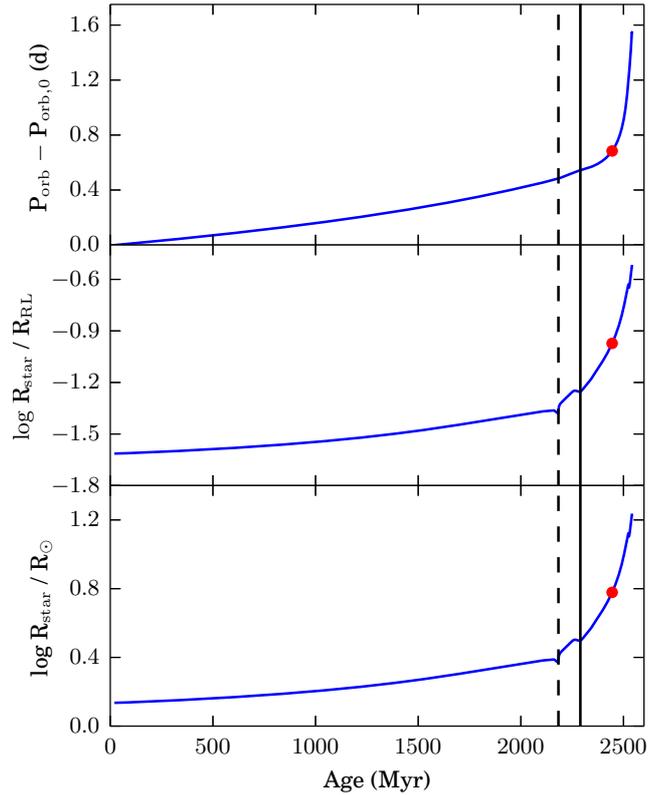}
\caption{Modelling the orbital period (top panel), Roche lobe filling factor (middle panel) and radius of the red giant main component (lower panel) of binary system with MESA with the approximated values of KIC\,5006817. The models start at the zero age main sequence (ZAMS) and stop with the onset of the RLOF. The dashed line marks the end of the hydrogen core burning (TAMS). The solid line marks the bottom of the RGB. The position of a model with the radius of 6\,$R$\sun, which equals the current state of KIC\,5006817 is shown as red dot.
\label{fig:binaryMesa} }
\end{figure}

Most hot subdwarf B (sdB) stars and cataclysmic variables (CV) are supposed to be produced from binaries that have undergone mass transfer and drastic mass loss during either a common envelope (CE) phase or a phase of stable Roche lobe overflow while on the RGB \citep{han2002,han2003,Nelemans2005}. Several unsolved questions remain regarding the exact sdB-progenitors and the details of the mass loss mechanism \citep{hu2008,ostensen2009,heber2009}.  

\subsection{Modelling the binary evolution of KIC\,5006817}
Currently, the binary system KIC\,5006817 is dynamically stable on timescales shorter than the evolutionary timescales of the red giant component. The system parameters, determined in the previous sections show that in its current configuration the orbit is wide compared to the Roche radii of the components. The low mass of the secondary component suggests that this star is most likely an M dwarf, which will stay on the main sequence for the remaining lifetime of the red giant companion. The evolution of the system will therefore mainly depend on the evolution of the red giant component (Fig.\,\ref{fig:binaryMesa}). 

MESA\footnote{Modules for Experiments in Stellar 
Astrophysics, see http://mesa.sourceforge.net} \citep{MESA2010, MESA2011, MESA2013}  is a one dimensional stellar 
evolution code originally designed for single star evolution. The code is adapted to handle binary systems by letting them evolve one step at a time, after which the orbital evolution is updated as well. The standard binary evolution that is included in MESA (version: November 2012) only handles the evolution of circular binaries. To check the evolution of an eccentric binary in which one of the components is a red giant, we implemented the equations derived by \cite{Verbunt95} to account for tidal interactions. The evolution of the eccentricity is governed by,
\begin{equation}
\frac{d \ln e}{dt} = -1.7 f \left(\frac{T_{\rm eff}}{4500 K}\right)^\frac{4}{3} \left(\frac{M_{\rm env}}{M_{\odot}}\right)^\frac{2}{3} \frac{M_{\odot}}{M} \frac{M_2}{M} \frac{M + M_2}{M} \left(\frac{R}{a}\right)^8 \frac{1}{yr}
\end{equation}
where $M_{\rm env}$ is the mass of the convective envelope of the RG star. $M_2$ is the mass of the companion and $a$ refers the semi-major axis of the orbit. The factor $f$ is calculated based on the mixing length parameter $\alpha$,
\begin{equation}
f = 1.01 \left(\frac{\alpha}{2}\right)^{(4/3)} .
\end{equation}
At every time step of MESA, $M_{\rm env}$ and $T_{\rm eff}$ are obtained from the stellar structure model. The start of the Roche lobe overflow (RLOF) is calculated based on the Roche-Lobe calculated at periastron passage.

To study the changes of eccentricity, period and the radius of the Roche lobe on such an eccentric binary system caused by the red giant evolution, we created an approximate model of the system KIC\,5006817, \new{using} the starting masses M$_{\rm RG} = 1.5\,M_{\odot}$ and M$_{\rm{MS}} = 0.3\,M_{\odot}$, an initial orbital period P$_{\rm{orbit},0} = 94.0$\,d and eccentricity to $e=0.70$. On the red giant branch a Reimers wind \citep{Kudritzki1978} was assumed, with \hbox{$\eta$=0.5}.

When we let this binary model evolve, we find that the orbital period slowly increases during the main sequence evolution (Fig.\,\ref{fig:binaryMesa}, top panel); 
\new{the rate of increase will accelerate as the star starts to ascend the red giant branch, mainly caused by mass loss due to the stellar wind on the RGB \citep{Hurley2002}.} The eccentricity remains stable. Only when the red giant is close to filling its Roche lobe, the system will circularize on the order of a few 10\,000 years. 

The middle panel of Fig.\,\ref{fig:binaryMesa} depicts how the expanding radius of the red giant component gradually grows to fill its Roche lobe \new{(R$_{\rm star}$/R$_{\rm Roche\,Lobe}<$1).} 
From the bottom panel of Fig.\,\ref{fig:binaryMesa}, it becomes obvious that the onset of RLOF at periastron passages will happen well below the tip of the RGB. \new{From MESA single models, we find that a giant of 1.5\,$M$\sun\ will have a maximum radius of about 140\,$R$\sun~before igniting helium.} Exactly how the orbital period will change during this process depends on how much of the transferred mass is accreted by the companion, and how much escapes the system. 

\begin{figure}[t!]
\includegraphics[width=0.49\textwidth, height=50mm ]{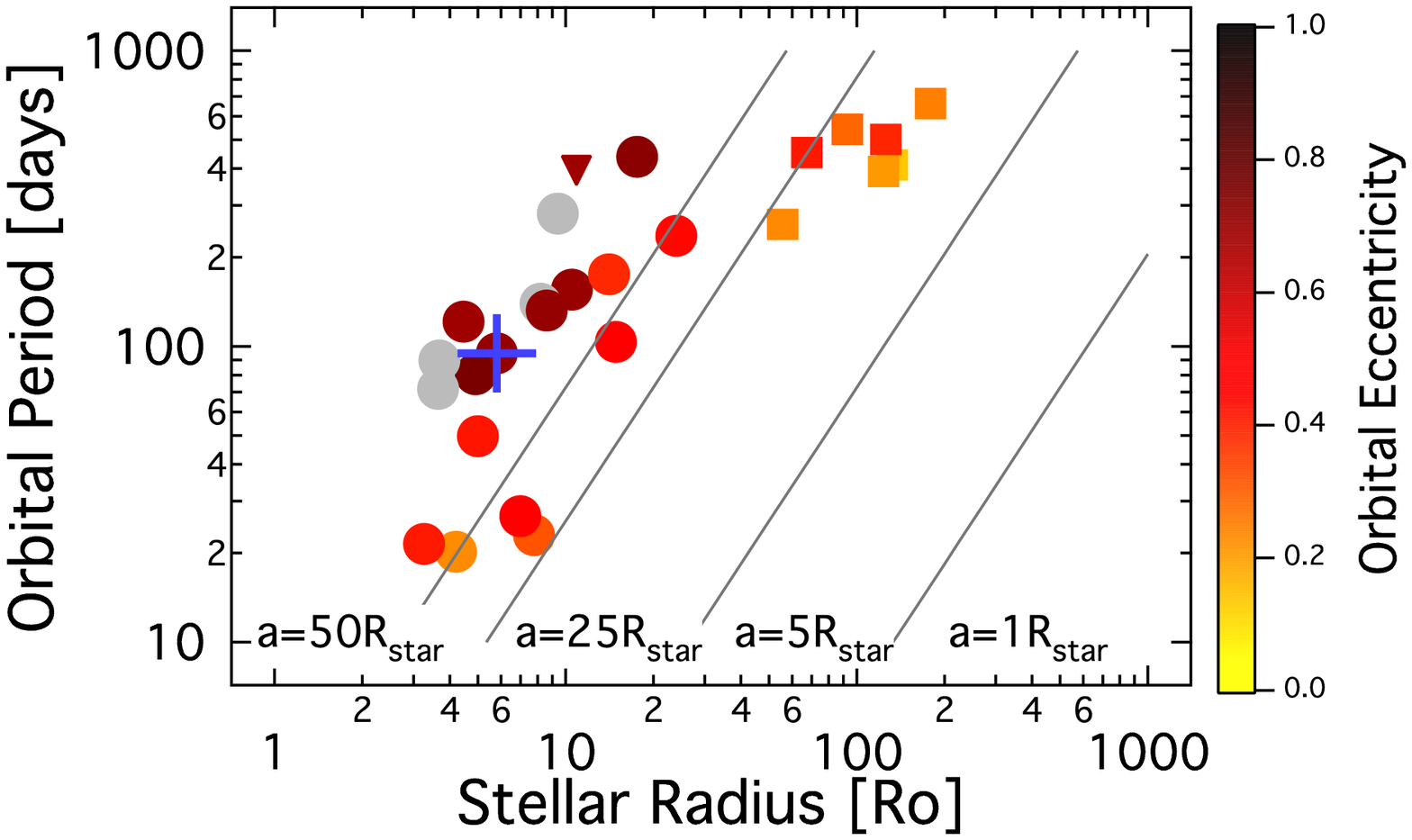}
\caption{\label{fig:rghbRadiusPeriod}
Stellar radius of the main component versus the orbital period of the binary system. Colour indicates the eccentricity of the systems (Table\,\ref{tab:sampleOrbits}). Stars from our sample (Tables\,\ref{tab:asteroseismicValues} and \ref{tab:shortCadenceGiants}) are plotted as filled dots. KIC\,5006817 is marked with a cross. The eclipsing red giant KIC\,8410637 \citep{hekker2010, frandsen2013} is shown as a triangle. The stars found in the LMC \citep{nicholls12} are shown as squares. Solid lines show the period of a system with a total mass of 1.8\,$M$\sun~and a semi-major axis, which is a multiple of the stellar radius. }
\end{figure}

\new{The periastron distance is 32$\pm1\,R_\odot$, which corresponds to $\sim$4 times the current radius of the red giant. When exceeding a radius, larger than the current semi-major axis of 106\,$R$\sun~(Table\,\ref{tab:ParamBestFitModel}), the red giant will completely engulf its companion in all phases of the orbit, leading to a CE phase while it is ascending the RGB.} \new{The interaction between two components during the CE phase for a rather similar system has been studied by \cite{han2002}, showing that the secondary component will spiral down inside the red giant primary.} \new{This phase happens rapidly compared to the evolutionary time scale of the red giant.}
A critical point is reached when the companion arrives at the bottom of the convective envelope and penetrates the initially radiative zone below it. The dynamical instability that follows will eject the envelope unless the companion is disrupted first \new{\citep{han2002}}. 

\new{Once the red giant ejects its envelope, we are left with the ``naked'' stellar core. If the red giant enters a CE-phase near the tip of the RGB, we would find a helium core which will become sdB stars. Otherwise, the system is likely to become a CV.}
Currently a majority of the observed long period sdB binaries have an eccentric orbit \citep{Vos2013}, indicating that their progenitor must have had a high eccentricity as well, as is the case with hearthbeat stars.
We therefore expect to find nearly exclusively stars in the H-shell burning phase in these eccentric systems and hardly any stars which have experienced He-core ignition and now settled on the red clump. The latter is only expected if the system's separation is large enough.

\begin{figure}[t!]
\includegraphics[width=0.49\textwidth, height=50mm]{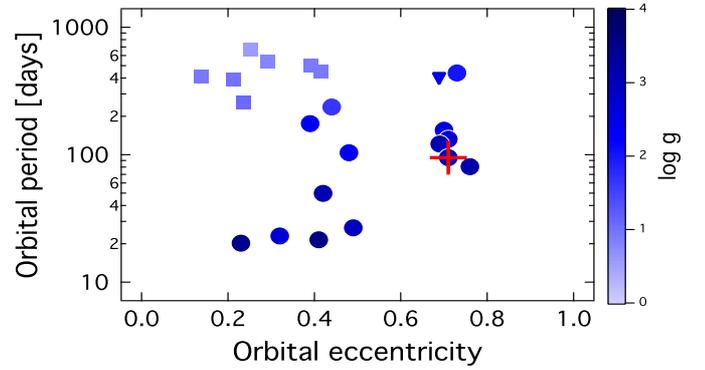}
\caption{Comparing the eccentricities of known eccentric systems with ellipsoidal modulation. Colour indicates the $\log g$ of the red giant component. Dots represent the stars from our sample for which we could derive the eccentricities from radial velocities. KIC\,5006817 and KIC\,8410637 are shown as a cross and triangle, respectively. Squares show the derived eccentricities of the stars in the LMC.  \label{fig:rghbPeriodEccentricity} }
\end{figure}

\subsection{Comparison to other known red giant stars in eccentric binary 
systems}

As a consequence of the expanding radius of the red giant, a relation between the radius of the primary and the orbital period \new{for such interacting systems} is expected, since only binaries with large enough periastron distances can stably exist. From seismology, we \new{obtained} the radii of a sample of red giant stars, located in such eccentric systems that are undergoing gravitational distortion, to a range between 2 and 24\,$R$\sun, as given in Table\,\ref{tab:asteroseismicValues}. \cite{nicholls12} reported seven similar systems from OGLE observations of the Large Magellanic Cloud (LMC). Although the OGLE lightcurves are not suitable for seismic investigations, they can constrain the fundamental parameters of the stars from modelling the phase diagram and photometric calibrations. The red giant systems found in the LMC are higher up in the HR~Diagram than the stars in our sample and also have in general longer orbital periods ($260-660$\,d). Fig.\,\ref{fig:rghbRadiusPeriod}, in which the stellar radii and orbital periods of the stars in the LMC and \textit{Kepler} sample are compared, shows that systems hosting \new{bigger} stars tend to have longer periods and ellipsoidal variation occurs for a specific combination of stellar radius and orbital period.
We note that our sample is biased due to the selection of stars which show the heartbeat effect. 
%
%
However, as shorter periods lead to smaller systems, which cannot stay stable for \newer{a long time.  These stars show a minimum orbital periods. In systems with shorter period mass transfer should eventually occur}.

In our ongoing spectroscopic campaign we have currently obtained enough radial velocity measurements for 14 systems in our sample to derive their orbital parameters. The full orbital parameters of these systems are given in Table\,\ref{tab:sampleOrbits}. 
Fig.\,\ref{fig:rghbPeriodEccentricity} depicts the eccentricities of the red giant heartbeat stars on the lower RGB found with \textit{Kepler} from our sample with the eccentricities of those close to the tip of the RGB \citep{nicholls12}.
The majority of the systems in Table\,\ref{tab:asteroseismicValues} for which we have RV data have eccentricities between $0.23 < e < 0.5$, which is compatible with the range of eccentricities reported by \cite{thompson2012} for their objects. Also, the 4 stars with periods shorter than 30\,d belong to the least eccentric in our sample. Among the long periodic systems, 5 exhibit eccentricities larger than 0.7. Our most eccentric system is KIC\,8144355 (P$_{\rm Orbit}$=80.6\,d, $e$=0.76$\pm$0.01).  

The systems found in the LMC so far have eccentricities that range between $0.15 < e < 0.45$, which suggests that the stars  might have experienced some circularization. It is not yet clear if mass transfer has already started in these systems or if the circularization is forced by tidal interaction. Yet the two samples are too small to draw a firm conclusion. In subsequent work, we will further investigate the distribution of eccentricities in the \textit{Kepler} red giant heartbeat stars from photometric and radial velocity data on larger datasets.

Finally, we note that KIC\,8410637, an eclipsing binary with a 408\,d period  \citep{hekker2010, frandsen2013} has a period and a high eccentricity (e$\sim$0.7), compatible with the other red giant heartbeat stars (Figs.\,\ref{fig:rghbRadiusPeriod}\,\&\,\ref{fig:rghbPeriodEccentricity}) but no heartbeat events are visible. 

\section{Conclusions}
In this \new{work} we have studied a sample of 18 red giant stars in eccentric binary systems, detected with the \textit{Kepler} satellite, that exhibit flux modulation as a result of \new{binary interaction} during their periastron passage. All giants in the systems in our sample exhibit solar-like oscillations, so we applied asteroseismic techniques to determine their global properties, as well as their evolutionary states. As shown in the analysis of one selected system, the approach of combining asteroseismic and binary modelling analyses is very powerful. 

\new{For KIC\,5006817, this new approach revealed a low mass companion with a mass of $M_2 = 0.29 \pm 0.03 \Msun$ (where the uncertainty encompasses both the beaming and non-beaming cases)}. Estimates based on the radial velocity curve and the optical spectrum  revealed that Doppler beaming should be contributing 300\,ppm to the light curve although the light curve modelling did not support this. This is possibly a consequence the long term trends in the \textit{Kepler} data, given that the period of KIC\,5006817 is $\sim$95\,d, which is very close to the length of a \textit{Kepler} quarter (90\,d).

\new{Through modelling the binary characteristics, while fixing the primary mass and radius to the asteroseismically determined values, the gravity darkening value was determined to be {\sc grd}\,=\,1.0\,$\pm$\,0.03 for the non-beaming case and {\sc grd}\,=\,$1.07 \pm 0.03$ for the beaming case. These values are inconsistent with the empirical value determined by \citet{Lucy1967}, {\sc grd}\,=\,0.32, and the more recent model dependent value determined by \citet{ClaretBloemen2011}, {\sc grd}\,=\,0.5. To obtain a binary model with values closer to those predicted by theory, the density of the primary component would need to deviate significantly from the well constrained asteroseismic value. For this reason we speculate that the gravity darkening values require further revision.} 

From modelling the binary evolution of an approximate binary system, we conclude that the system is in fact too young to be synchronized. When comparing the properties of the full sample (Table\,\ref{tab:asteroseismicValues}), we found a correlation between the radius of the primary red giant component and the orbital period. Furthermore, all stars show seismic characteristics of stars in the state of H-shell burning. For a few, we cannot rule out a membership of the red clump or AGB. We argue that this is an effect of stellar evolution as the red giant's radius along the red giant branch will increase until the helium core ignites. If a system gets too close, it can undergo a common envelope phase which could lead to the ejection of the convective envelope of the red giant. This scenario is a potential evolutionary channel for the formation of cataclysmic variables and sdB stars. 

Our sample is an interesting class of ellipsoidal variables which offers unique conditions to study interactions in and the evolution of eccentric binary systems.
\new{Optimal cases are systems which show the heartbeat effect and exhibit primary and secondary eclipses or are double lined spectroscopic binaries, as for such systems also for the secondary component independent fundamental parameters can be derived.}
The ensemble of stars presented here allows us to study the binary interaction and the future fate of such eccentric systems in a new way, to help unravel common-envelope physics. 

\begin{acknowledgements}
We acknowledge the work of the team behind \textit{Kepler}. Funding for the \textit{Kepler} Mission is provided by NASA's Science Mission Directorate. The groundbased follow-up observations are based on spectroscopy made with the Mercator Telescope, operated on the island of La Palma by the Flemish Community, at the Spanish Observatorio del Roque de los Muchachos of the Instituto de Astrof'sica de Canarias. This work partially used data analyzed under the NASA grant NNX12AE17G. This research is (partially) funded by the Research Council of the KU Leuven under grant agreement GOA/2013/012.
The research leading to these results has received funding from the European Research Council under the European Community's Seventh Framework Programme (FP7/2007--2013)/ERC grant agreement n$^\circ$227224 (PROSPERITY). 
The research leading to these results has received funding from the European Community's Seventh Framework Programme FP7-SPACE-2011-1, project number 312844
(SPACEINN).
K.H. was supported by a UK STFC PhD grant. JDR, T.K. and E.C. acknowledge the support of the FWO-Flanders under project O6260 - G.0728.11. 
TK also acknowledges financial support from the Austrian Science Fund (FWF P23608).
A.T. was supported by the Fund for Scientific Research. S.H. was supported by the Netherlands Organisation for Scientific Research (NWO). 
V.S.S. is an Aspirant PhD Fellow of the FWO, Belgium. AD is supported by a J\'anos Bolyai Research Scholarship of the Hungarian Academy of Sciences. This project has been supported by the Hungarian OTKA Grants K76816, K83790, MB08C 81013 and KTIA URKUT\_10-1-2011-0019 grant and the ``Lend\"ulet-2009'' Young Researchers Program of the Hungarian Academy of
Sciences.%
P.G.B thanks Nick Cox for observational work. 

\end{acknowledgements}

\bibliographystyle{aa}
\bibliography{rotationReferences}

\end{document}